\documentclass[11pt,a4paper]{article}
\pdfoutput=1 
\usepackage{jcappub}
\usepackage{graphicx}
\usepackage[utf8x]{inputenc}
\usepackage[T1]{fontenc}
\usepackage{subeqnarray}
\DeclareMathOperator\arctanh{atanh}
\DeclareMathOperator\arcsinh{asinh}
\newcommand\apj{\ref@jnl{ApJ}}

\def\gmn{g_{\mu\nu}}
\def\om{\omega}
\def\covD{\nabla}
\def\la{\lambda}
\def\La{\Lambda}
\def\al{{\alpha^2}}
\def\alfo{\frac{1}{\alpha}}

\title{Galileons and strong gravity }
\author[a,b]{Javier Chagoya,}
\author[a]{Kazuya Koyama,}

\author[b]{Gustavo Niz}

\author[a]{and Gianmassimo Tasinato}
\affiliation[a]{
Institute of Cosmology \&  Gravitation, University of Portsmouth, Dennis Sciama Building, Portsmouth, PO1 3FX, United Kingdom
}
\affiliation[b]{
Departamento de F\'isica, Universidad de Guanajuato,\\
DCI, Campus Le\'on, C.P. 37150, Le\'on, Guanajuato, M\'exico.\\
}
 \emailAdd{jfchagoya@fisica.ugto.mx}
\emailAdd{kazuya.koyama@port.ac.uk}
\emailAdd{g.niz@ugto.mx}
\emailAdd{gianmassimo.tasinato@port.ac.uk}

\abstract{
In the context of a cubic Galileon model in which the Vainshtein mechanism suppresses the scalar field interactions with matter, we study low-density stars with slow rotation and static relativistic stars. We develop an expansion scheme to find approximated solutions inside the Vainshtein radius, and show that deviations from General Relativity (GR), while considering rotation, are also suppressed by the Vainshtein mechanism. In a quadratic coupling model, in which the scalarisation effect can significantly enhance deviations from GR in normal scalar tensor gravity, the Galileon term successfully suppresses the large deviations away from GR. Moreover, using a realistic equation of state, we construct solutions for a relativistic star, and show that 
deviations from GR are more suppressed for higher density objects. However, we found that the scalar field solution ceases to exist above a critical density,
which roughly corresponds to the maximum mass of a neutron star. This indicates that, for a compact object described by a polytropic equation of state, 
the configuration that would collapse into a black hole cannot support a non-trivial scalar field. }

\begin{document}
\maketitle
\flushbottom
 \section{Introduction}
There is strong evidence that the present Universe is expanding in an accelerated way \cite{Perlmutter:1998np}, and the standard model of cosmology presumes that a cosmological constant is responsible for this acceleration. The vacuum energy of the Universe would contribute to this cosmological constant, but a theoretical prediction for the vacuum energy is several orders of magnitude larger than the observed value to explain the acceleration. This theoretical complication has lead to alternative scenarios in order to explain the late time acceleration of the Universe, and also to solve the vacuum energy problem. As explained by Weinberg \cite{Weinberg:1988cp}, one has to modify General Relativity (GR) in order to relax dynamically the value of the vacuum energy to smaller values. Following this line of thought, several infrared modified gravity theories have been proposed which, in principle, could achieve self-acceleration in the Universe due to gravitational degrees of freedom. However, there are 
stringent bounds on 
gravitational interactions which are different from GR in local (solar system) experiments. Therefore, for these alternatives to be physical, it is necessary to include a screening mechanism that ``hides'' the effects of such  modifications to GR (see \cite{Joyce:2014kja} for a review and references therein). One interesting mechanism, proposed by Vainshtein in 1972 \cite{Vainshtein}, hides GR modifications due to non-linear derivative self-couplings. This mechanism can be embedded in the Galileon models \cite{Nicolis:2008in, Deffayet:2009wt}. 

The Galileon Lagrangian was initially introduced by Horndeski \cite{Horndeski:1974wa} few decades ago, as a subset of  the most general scalar-tensor theory with second order equations of motion. This was rediscovered more recently by Nicolis et al \cite{Nicolis:2008in}. The Galileon Lagrangian has emerged as the effective theory of some theories  such as the DGP model \cite{Dvali:2000hr}, the recently proposed ghost-free massive gravity \cite{deRham:2010ik}, and   vector theories that break
 in a specific way an Abelian symmetry \cite{Tasinato:2014eka}.  Either within the embedding modified gravity theories or the simple Galileon Lagrangians, the Vainshtein mechanism has been intensively studied for static and spherically symmetric backgrounds in the weak field limit (see for example \cite{vainshteinpast, Babichev:vainshtein, us, Barreira:2013eea}). However, little work has been done away from staticity and spherical symmetry, such as in two body systems \cite{binaries} or cosmology \cite{vainshteincosmo, Barreira:2013eea}. In this work we try to give some insights into the implications of renouncing to 
staticity and to weak gravity approximations. For this purpose, we will consider the simplest Galileon model, which only includes the cubic Galileon term and find approximated solutions for slowly rotating low-density stars initially, and relativistic stars towards the end.

The paper is organised as follows. In Section II, we present the model and the equations for spherically symmetric vacuum configurations. We first review the solution for a scalar field in the Minkowski background and develop a perturbation scheme that works inside the Vainshtein radius. We apply this method to find first order corrections to the vacuum solutions in GR due to the Galileon term. We also discuss the effect of adding rotation. In Section III, we introduce a coupling to matter and discuss how the solution with a constant density source matches to these vacuum solutions. In section IV, we study the scalarisation phenomenon (strong amplification of the effective scalar-matter coupling) in a quadratic coupling model, and study whether this happens in the presence of the cubic Galileon term. We then study more realistic neutron stars and investigate how the Vainshtein mechanism suppresses the modification of gravity in the strong gravity regime. We also discuss the condition for existence of the 
scalar field solution.

\section{Minimally coupled cubic Galileon action and equations of motion}
Our starting point is the Einstein-Hilbert action, minimally coupled to a scalar field with the cubic Galileon term with no additional
matter included
\begin{equation}
S=\int d^4 x \sqrt{-g}  \left( M_{p}^2 R - \frac{{1}}{2}g^{\mu\nu}\partial_\mu\Phi\partial_\nu\Phi-\frac{\al}{\Lambda^3}\Box\Phi g^{\mu\nu}\partial_\mu\Phi\partial_\nu\Phi\right), \label{action}
\end{equation}
where $\alpha$ is a dimensionless constant, $M_{p}$ is the reduced Planck mass and $\Lambda$ is a constant with dimensions of mass.
In models where the cubic Galileon term is associated with the late time acceleration of the Universe, $\Lambda^3 \alpha^{-2}$ is typically given by ($1000$ km)$^{-3}$. Varying this action with respect to the metric and the scalar field, we obtain the following equations of motion:
\begin{subeqnarray}
\xi_{\mu\nu}&:=&M_{p}^2\left(R_{\mu\nu}-\frac{1}{2}g_{\mu\nu}R\right)=  \partial_\mu\Phi\partial_\nu\Phi-\frac{1}{2} \gmn\partial^\alpha\Phi\partial_\alpha\Phi\nonumber \\
   & &  + \frac{\al}{\Lambda^3}\left(\Box\Phi\partial_\mu\Phi\partial_\nu\Phi-2\partial_\mu\Phi(\covD_\nu\covD_\alpha\Phi)\partial^{\alpha}\Phi + \gmn\partial_\alpha\Phi(\covD^{\alpha}\covD^{\beta}\Phi)\partial_{\beta}\Phi\right),\slabel{em1}
\\
\xi_\Phi&:=& \Box\Phi+\frac{2\al}{\Lambda^3}\left[- \covD^\beta\covD^\nu\Phi \covD_\beta\covD_\nu\Phi+ (\Box\Phi)^2-  R^{\mu\nu}\partial_\mu\Phi\partial_\nu\Phi\right]=0\slabel{sc1}.\label{eqsmov}
\end{subeqnarray}
One can understand the appearance of the Ricci tensor in \eqref{sc1} from the commutation of the third order derivatives of the scalar field $[\covD_\alpha,\covD_\nu]\covD_\beta\Phi=R^\rho{}_{\beta\alpha\nu}\covD_\rho\Phi$.  

Let us consider a static and spherically symmetric spacetime, together with a scalar field that depends on the radial coordinate only, namely
\begin{subeqnarray}
ds^2&=&-e^{2\nu(r)}dt^2+e^{2\lambda(r)}dr^2+ r^2 (d\theta^2+\sin^2\theta d\varphi^2),\nonumber\\
\Phi&=&\Phi(r).
\end{subeqnarray}
Later on we will depart from this assumption by considering a slowly rotating spacetime (with an off-diagonal $dtd\varphi$ term).
Under the previous assumption, the equations of motion \eqref{eqsmov} then reduce to
\begin{subeqnarray}\label{eqso}
0= \xi^0{}_0&=&M_{p}^2 \left[-4 e^{2 \lambda{}} \left(-1+e^{2 \lambda{}}\right) \Lambda ^3-8 e^{2 \lambda{}} r \Lambda ^3 \lambda'{}\right]-4 \al r^2 \lambda'{} {\Phi}'{}^3 \nonumber \\
& &-r^2 {\Phi}'{}^2 \left(-{} e^{2 \lambda{}} \Lambda ^3+4 \al {\Phi}''{}\right),\\ 
0=\xi^1{}_1&=&-e^{2 \lambda{}} r^2 {} \Lambda ^3 {\Phi}'{}^2-4 e^{2 \lambda{}} M_{p}^2 \Lambda ^3 \left(-1+e^{2 \lambda{}}-2 r \nu'{}\right)-4 \al r {\Phi}'{}^3 \left(2+r \nu'{}\right),\\ 
0=\xi^2{}_{2}&=&4 \al r^2 \lambda'{} {\Phi}'{}^3-r^2 {\Phi}'{}^2 \left(-e^{2 \lambda{}} {} \Lambda ^3+4 \al {\Phi}''{}\right)\nonumber  \\
& & +{M_p}^2 r \left[-4 e^{2 \lambda{}} \Lambda ^3 \lambda'{} \left(1+r \nu'{}\right)+4 e^{2 \lambda{}} \Lambda ^3 \left(\nu'{}+r \nu'{}^2+r \nu''{}\right)\right],\\ \label{meteq}
0=\xi_\Phi&=&e^{2 \lambda{}} {} \lambda'{} {\Phi}'{}+e^{2 \lambda{}} r  {\Phi}'{} \left(-2-r \nu'{}\right)-e^{2 \lambda{}} {} {\Phi}''{} +\frac{\al}{r^2  \Lambda ^3} \left[4 r {\Phi}'{} \left(-2-r \nu'{}\right) {\Phi}''{}\right.\nonumber \\
& &\left.+2 {\Phi}'{}^2 \left(-2-4 r \nu'{}-r^2 \nu'{}^2+3 r \lambda'{} \left(2+r \nu'{}\right)-r^2 \nu''{}\right)\right].
\end{subeqnarray}
The equation for the scalar field \eqref{meteq} can be integrated once, as can be expected from the fact 
that the Lagrangian does not depend on $\Phi$ itself but on its derivatives only. The integration of this equation results in
\begin{equation}
e^{\nu-\la}r^2{}\La^3\Phi' + 2\al r e^{\nu-3\la}(2+r\nu')\Phi'{}^2  =\La^3  \zeta \label{zeta},
\end{equation}
where the dimensionless integration constant $\zeta$ must be determined by boundary conditions. 
This is an algebraic equation for $\Phi'$, with solutions
\begin{equation}
\Phi' = \frac{-e^{\nu-\la} r^2 {} \La^3 \pm \sqrt{e^{2(\nu-\la)} r^4  \La^6 + 8\al\La^3 \zeta r e^{\nu - 3\la} (2+ r\nu')}}{4\al r e^{\nu - 3\la} (2 + r\nu')}. \label{solquad}
\end{equation}
From this  expression, one notices that in the limit of large $\alpha$ (keeping all other quantities fixed)
$\Phi'$ scales as $1/\alpha$, plus subleading corrections. When substituting the scalar field solution into the Einstein equations, one finds that the Galileon contributions are suppressed by  powers of $1/\alpha$, hence they can be neglected in the large $\alpha$ limit. This result suggests that we can analyse
this Einstein-Galileon system, in the large $\alpha$ limit, by a perturbative expansion in powers of $1/\alpha$. This is the approach we will follow.

%%%%%%%%%%%%%%%%%%%%%%%%%%%%%%%%%%%%%%%%%
\subsection{A test scalar field }
%%%%%%%%%%%%%%%%%%%%%%%%%%%%%%%%%%%%%%%%%%
Before studying the full system, it is 
illustrative to look at the behaviour of a test scalar field on a fixed background. If we consider the Minkowski spacetime as our background, the solution for $\Phi'$ reduces to
 \begin{equation}
\Phi'_{flat}=-\frac{r {} \Lambda ^3}{8 \al}+\frac{\Lambda ^{3/2}}{8\al} \sqrt{\frac{16 \al {\zeta}}{r}+r^2 \Lambda ^3}\label{flat},
\end{equation}
where we have chosen the positive sign in \eqref{solquad} so that, in the limit $r\to \infty$, we recover $\Phi'_{flat}=0$ modulus terms with $\mathcal O(r^{-2})$. 
In the large $\alpha$ limit, we can expand the previous solution to obtain
\begin{equation}
\Phi'_{flat} = \left(\frac{\Lambda^{3}{\zeta}}{4 {r} {\al }}\right)^{1/2}-\frac{\Lambda^3 r {} }{8 \al }+\left( \frac{\Lambda^{9}r^{5}}{2^{12} {\zeta} \alpha ^{6}}\right)^{1/2} +O\left({\alpha }^{-4}\right) .\label{phiexp1}
\end{equation}
 This expansion is reliable only if the magnitude of every successive
term is smaller than the previous one. Since the second term in the expansion arises from the kinetic term, the condition that the first term to dominate over the second term is equivalent to the condition that the contribution from the Galileon term is dominant 
 with respect to   the canonical kinetic term. 
The equality of the first two terms defines the `Vainsthein' radius,
\begin{equation}
r_V^3 = \frac{16\al \zeta}{\La^3 } \label{rcrit}.
\end{equation}
For $r<r_V$, both the validity of the series expansion and the 
dominance of the Galileon term over the kinetic term are guaranteed. Furthermore, the ratio of the solution \eqref{phiexp1} to that without the Galileon term ($\Phi_{can}'\sim r^{-2}$) is given by 
\begin{equation*}
\frac{\Phi'_{flat}}{\Phi'_{can}}\sim\left(\frac{{\al}}{r^{3}\La^{3}}\right)^{1/2}\sim\left(\frac{r_V}{r}\right)^{3/2}.
\end{equation*}  
This suggests that \eqref{phiexp1} can be written as
\begin{equation}
\phi'_{flat}\sim\frac{1}{r^2}\left[ \left(\frac{r}{r_V}\right)^{3/2} + 
\left(\frac{r}{r_V}\right)^3 + \left(\frac{r}{r_V}\right)^{9/2}  + \dots\right].\label{appdual}
\end{equation}
Thus the $\alpha^{-1}$ expansion is equivalent to the expansion in terms of 
$(r/r_V)^{3/2}$ in this case. In the following, we will apply this expansion scheme to find a solution deep inside the Vainshtein radius $r<r_V$. It was argued that in this regime, there exists a dual description of the theory that admits perturbative solutions \cite{Gabadadze2012,Padilla:2012ry}. If applied to a test scalar field in the Minkowski spacetime, the perturbative expansion in the dual formulation is controlled by $(r/r_V)^{3/2}$, and thus agrees with our expansion. 
 
Although we will study the coupling to matter in the next section, in order to gain some insights into the integration constant $\zeta$, it is worthwhile a brief digression to look at the scalar field solution in the presence 
of a test point source, coupled to the scalar field around a flat background. The point particle may be introduced in the original Lagrangian by adding the mass term
$$\mathcal L_{point} =- \frac{1}{2}\beta\frac{\Phi }{M_{p}}{T},$$ 
%- 2\frac{\hbar}{c^3}\frac{\beta}{M_{p}}\Phi \frac{T}{4},$$ 
with $T=M \delta^{(3)}(r)$.
The scalar field equation then becomes
\begin{align}
%\frac{1}{r^2} \frac{d}{dr}\left[r^2{}\Phi' + 4\frac{\alpha}{\La^3} r\Phi'{}^2   \right]& =  \frac{\beta M_{\odot}}{2 M_{p}}\delta^{(3)}(r) 
%= \frac{ \hbar}{8\pi c} \beta\frac{M_{\odot}}{  M_{p}}\frac{\delta(r)}{4\pi r^2}
%, \nonumber \\
%\Rightarrow\hspace{3em}
\frac{\Phi'}{r} + \frac{4\al}{{} \La^3}\left(\frac{\Phi'}{r}\right)^2 &= \frac{\beta}{8\pi r^3} \frac{M}{  M_{p}}, \label{rv}
\end{align}
which is the same as equation \eqref{zeta} in a flat background, but with the following identification of variables
\begin{equation}
\zeta=\frac{\beta  M}{8\pi   M_{p}}. \label{zetadef}
\end{equation}
As a conclusion, the scalar field charge, which is proportional to $\zeta$, plays essentially the same role as a mass. To close this digression, we also note that under this identification of $\zeta$, the Vainshtein radius $r_V$ reads
\begin{equation} 
r_V^3 =   \frac{2\al\beta M }{\pi M_{p} \La^3 },
\label{vr}
\end{equation}
which agrees with the expressions in the literature, up to a redefinition of $\beta$, for the Vainshtein radius (see for example \cite{Kaloper:2011qc}).

%%%%%%%%%%%%%%%%%%%%%%%%%%%%%%%%%%%%%%%%%%%%%
\subsection{Large $\alpha$ expansion}
%%%%%%%%%%%%%%%%%%%%%%%%%%%%%%%%%%%%%%%%%%%%%
We now turn our attention to the study of the solutions in the full system using the large $\alpha$ expansion. At the leading order, we expect to recover GR solutions due to the Vainshtein mechanism. For a spherically symmetric system in GR, the solution is uniquely determined by the Schwarzschild metric. However, there is no Birkhoff's theorem in our system thus we cannot exclude a possibility to have solutions different from Schwarzschild. Here we are interested in a solution which is close to Schwarzschild within the Vainshtein radius to recover GR, hence we would like to see if it can can be consistently constructed using our large $\alpha$ expansion. 

At the leading order, this is equivalent to consider a test Galileon field, which accretes into the Schwarzschild black hole, as it was done in \cite{babichevSchw}. Then, one could analyse the backreaction of such a field  (see \cite{babichevRev}). Here, we will calculate the metric corrections using our large $\alpha$ approximation. This approach should give some insights into the Vainshtein mechanism under strong gravitational fields. Unlike \cite{babichevSchw, babichevRev} where the existence of a black hole surrounded by a Galileon scalar field is discussed in the presence of non-trivial cosmological boundary conditions, here we are interested in finding the interior and exterior solutions of a massive body in an asymptotically flat and static spacetime. Therefore, we begin by constructing the vacuum exterior solution in this section. Later on, we will study the interior solution, and also discuss what happens when the mass of such a configuration approaches the limit where it might collapse and form a 
black hole. 

In \cite{Kaloper:2011qc}, the authors offered a demonstration of the Vainshtein mechanism around a thin-shell configuration of matter, showing that asymptotically flat solutions exist. Here, under the large $\alpha$ approximation, we construct explicitly solutions around a static matter configuration, and confirm that below the Vainshtein radius the behaviour of the scalar field is determined by the Galileon term. We also provide the explicit matching between the interior solution for a static source of constant density and the exterior Galileon solution. The transition from this solution to the asymptotically flat solution beyond $r_V$ is not done explicitly, but it is guaranteed by the sign choice made for $\Phi$.

Our method is a generalization of what we did for a test scalar field in  Minkowski spacetime. First we solve for the scalar field on a Schwarzschild background metric and then we look for the first order corrections to the metric in the $\alpha^{-1}$ expansion. We expand our functions in 
powers of $\alpha^{-1}$ as
\begin{subeqnarray} \label{seriesone}
\nu(r)&=&\nu_0(r) + \frac{1}{\alpha}\nu_1(r) +\mathcal O(\alpha^{-2})\simeq \frac{1}{2} \log\left(1- \frac{r_s}{r}\right)+ \frac{1}{\alpha}\nu_1(r),\\
\lambda(r)&=&\lambda_0(r) + \frac{1}{\alpha}\lambda_1(r) )+\mathcal O(\alpha^{-2})\simeq \frac{1}{2} \log\left(1- \frac{r_s}{r}\right)^{-1}+ \frac{1}{\alpha}\la_1(r),\\
\Phi(r)&=&  \frac{1}{\alpha}\Phi_1(r)+  \frac{1}{\al}\Phi_2(r) +\mathcal O(\alpha^{-3}), \label{sfex}
\end{subeqnarray}
where $r_s$ is the Schwarzschild radius associated with a mass $M$.
The first term of the scalar field expansion, $\Phi_1$, is fully determined by $\nu_0$ and $\la_0$ in the scalar field equation.  
Altogether, $\nu_0$, $\la_0$ and $\Phi_1$ determine
$\nu_1$ and $\la_1$ via the terms of 
 order $\alpha^{-1}$  in the Einstein equations \eqref{eqso}. After inserting these functions in the
corresponding terms of the same order in the scalar field expansion \eqref{zeta}, one can solve for $\Phi_2$. Finally, we can identify the contribution from the canonical kinetic term in $\Phi$, and obtain an expression for the `Vainshtein radius' as the radius at which the Galileon contribution becomes comparable to the contribution from the canonical kinetic term. 

Either by expanding \eqref{solquad}, or by directly solving the scalar field equation \eqref{zeta} at the lowest order
in $\alpha^{-1}$, we find that the solution for $\Phi_1$ is
\begin{equation}
\Phi_1'{}^2=\frac{\zeta_0 r \Lambda^3}{(r-{r_s}) (4 r-3{r_s})}, \label{phischw}
\end{equation}
where $\zeta_0$ has the same interpretation as before: the scalar charge developed by $\Phi$ on a given background metric. In the limit $r_s\to 0$, equation \eqref{phischw} is exactly the $\mathcal O(\alpha^{-1})$ term of the flat solution \eqref{phiexp1}. To be completely general, we should rewrite the integrated scalar field equation \eqref{zeta} as
\begin{align}
\left[\prod_{i=0} e^{\frac{\nu_i-\la_i}{\alpha^{i}}}\right]r^2{}\La^3\sum_{i=1}\frac{\Phi'_i}{\alpha^{i}} + 2\al r \left[\prod_{i=0}e^{\frac{\nu_i-3\la_i}{\alpha^{i}}}\right](2+r\nu')\left[\sum_{i=1}\frac{\Phi'_i}{\alpha^{i}} \right]^2&  = \La^3\zeta\nonumber \\
& \equiv\La^3 \sum_{i=0} \frac{\zeta_i}{\alpha^{i}}, \label{phigen}
\end{align}
and solve this equation at each order in the $\alpha^{-1}$ expansion. This shows that the total scalar charge is still a conserved quantity but it is artificially split into parts coming from different orders in the expansion \eqref{seriesone}. For example, $\zeta_0$ in \eqref{phischw} has a contribution from the Schwarzschild part of the metric only.

We now insert $\nu_0, \la_0$ and $\Phi_1$ in the Einstein equations, obtaining two independent equations for $\la_1$ and $\nu_1$;
\begin{subeqnarray}
\frac{1}{M_{p}^2} \sqrt{\frac{\zeta_0^3\La^3r^3(r_s-r)}{(3r_s-4r)^5}}+\left[(r_s-r)\la_1\right]'&=&0,   \label{eqscorra}\\
\frac{1}{4 M_{p}^2}\sqrt{\frac{\zeta_0^3 r \La^3}{(3r_s-4r)(r_s-r)}}+\la_1 + (r_s-r)\nu_1'&=&0. \label{eqscorr}
\end{subeqnarray}
These equations can be solved exactly, giving us the dominant Galileon corrections to the Schwarzschild metric, which 
could then be used to explore strong gravity effects. The solutions have rather complicated expressions, hence we only display explicitly $\la_1$ but leave $\nu_1'$ unintegrated. The solutions are given by
\begin{subeqnarray} \label{metrico1}
\lambda_1 &=&\frac{ (\zeta_0\La)^{3/2}}{M_{p}^2}\frac{20 (r_s r)^{3/2}-3 \sqrt{r_s r}\left(3r_s^2 +4 r^2\right)+2 r_s  (4r-3r_s)^{3/2}(r-r_s)^{1/2} 
E_{\frac{1}{2}}\left[\arcsinh\sqrt{\frac{r_s}{r-r_s}}\right]}{24 (4 r-3 r_s)^{3/2} (r-r_s) \sqrt{r_s (r-r_s)}} \nonumber \\
& &+\frac{\lambda_{10}}{r_s-r}, \\
\nu_1'&=&-\frac{(9 r_s-10 r) (r_s-2 r) \left(\frac{r \zeta_0 \Lambda}{r-r_s}\right)^{3/2}}{4 M_{p}^2 r (4 r-3r_s)^{5/2}}-\lambda_1'(r), \label{nupeq}
\end{subeqnarray}
where $\lambda_{10}$ (and $\nu_{10}$ that we will meet next)
 are
  integration constants associated with
eqs. (\ref{eqscorra}), while  
$E_k[x]=\int_0^x\sqrt{1-k^2\sin^2\theta}d\theta$ is an elliptic integral of the second kind. 

For future use, we present the weak gravity limit ($r_s/r \ll 1$) for $\la_1$ and $\nu_1$,
\begin{eqnarray}\label{weakcorrectionsexterior}
\lambda_1& =& -\frac{({\zeta_0} \Lambda )^{3/2}}{16 M_{p}^2 {r^{1/2}}}+\frac{3  r_s({\zeta_0} \Lambda )^{3/2}}{128 M_{p}^2 r^{3/2}}+\frac{3 r_s^2({\zeta_0} \Lambda )^{3/2}}{128 M_{p}^2 r^{5/2}}-\frac{\lambda _{10}^{{}}}{r} -\frac{r_s\lambda _{10}^{{}}}{r^2}-\frac{r_s^2 \lambda _{10}^{{}}}{r^3},\\
\nu_1 &=&\frac{3 ({\zeta_0} \Lambda )^{3/2}}{8 M_{p}^2 {r^{1/2}}}+\frac{35  r_s({\zeta_0} \Lambda )^{3/2}}{192 M_{p}^2 r^{3/2}}+\frac{r_s\lambda _{10}^{{}}}{r^2}+\frac{\lambda _{10}^{{}}}{r}+\nu _{10}^{{}}.
\end{eqnarray}
Now we can find the solution for $\Phi_2$ using \eqref{phigen}. The resulting solution is exact, however the expression is too lengthy and it is sufficient to stress that the metric and scalar field solutions are all consistent with the Minkowski background solutions that we found in the previous section in the weak field limit $r_s/r\ll 1$. Moreover, in order to estimate the Vainshtein radius, we only need the contribution from the canonical kinetic term to 
$\Phi_2$, which is given by
\begin{equation}
-\frac{r}{8}\left(1-\frac{3}{4}\frac{r_s}{r}\right)^{-1}{}  \Lambda ^3. \label{kinpart}
\end{equation}
This expression should be compared to $\Phi_1$ in order to obtain the radius at which the Galileon cubic term becomes of the same order as the canonical kinetic term. In other words, the following equality should be satisfied
\begin{align}
\frac{1}{\al}\frac{\zeta_0 r \Lambda^3}{(r-{r_s}) (4 r-3{r_s})}&\sim \frac{1}{\alpha^4}\frac{r^2}{64}\left(1-\frac{3}{4}\frac{r_s}{r}\right)^{-2}  \Lambda ^6 , \nonumber \\
%\frac{1}{\alpha}\frac{(4 r-3{r_s})\zeta_0 r \Lambda^3}{(r-{r_s}) }&\gg\frac{1}{\alpha^2}\frac{r^4}{4}{}^2  \Lambda ^6  \nonumber \\
\end{align} 
which reduces to
\begin{align}
\frac{r-{r_s}}{ r-\frac{3}{4}{r_s}} r^3 &\sim \frac{16  \al \zeta_0}{\La^3}  .\label{vrs} 
\end{align} 
Notice that the Vainshtein radius in the Schwarzschild spacetime is always larger
than that in the flat background, \eqref{rcrit}, because the factor in front of $r^3$ is always less than one. This last claim however assumes that the integration constant $\zeta_0$ 
has the same value both on a Schwarzschild and on a Minkowski background. In reality they may differ because
they are given by matching \emph{different} scalar field solutions for each background to the
solutions inside the matter configuration under consideration. It
is important to stress that in order to derive this results we did not use any weak gravity approximation, but
only neglected contributions to $\Phi_2$ coming from $\lambda_1$ and $\nu_1$. However, from the equations \eqref{eqscorr} it is possible to deduce that these metric corrections are, at most, of order $\alpha^{-1}$, as in the Minkowski case. 

%%%%%%%%%%%%%%%%%%%%%%%%%%%%%%%%%%%%%%%%%%%%%
\subsection{Rotation}
%%%%%%%%%%%%%%%%%%%%%%%%%%%%%%%%%%%%%%%%%%%%%
In the previous section, we obtained static spherically symmetric solutions using our $\alpha^{-1}$ expansion without taking weak gravity limit. In this section, we will include rotation into the solution. It is  difficult to find solutions with fast rotations analytically, so we focus on a slow rotation component only.

Let us consider a slowly rotating spacetime characterized by a small function
$\omega(r)$. To first order in this function, the metric is just the same as in the static case but with a new $d\varphi \,dt$ 
component in the metric, 
\begin{equation}
ds^2=-e^{2\nu(r)}dt^2+e^{2\lambda(r)}dr^2+ r^2 (d\theta^2+ \sin^2\theta d\varphi^2)+2 r^2   \omega(r) \sin^2\theta d\varphi dt. \label{srotmetric}
\end{equation}
This additional metric component is determined by a new equation of motion, $\xi^\varphi{}_t$, 
which, because of $\partial_\varphi\Phi = 0$, does not have explicit contributions from the scalar field and is therefore the same as in GR. 
In vacuum this equation can be integrated once and the resulting integration constant is just the angular momentum,
\begin{align}
J = e^{-\la-\nu} r^4 \om'. \label{rotfull}
\end{align}
Expanding $\nu$ and $\la$ as in \eqref{seriesone}, it is then natural to split $\omega$ in an $\mathcal O(\alpha^0)$ part, which is just the usual Schwarzschild solution, and an $\mathcal O(\alpha^{-1})$ part, which has to be determined from the metric corrections $\nu_1$ and  $\la_1$ given by \eqref{metrico1}. Also, the total angular momentum $J$ is separated into contributions
coming from different orders in $\alpha$, $J=\sum_{n=0} {\alpha^{-n}}J_n$. Then
\begin{eqnarray}
\omega(r) &=&\omega_0(r)+\alfo\omega_1(r) + \dots \\
&=& -\frac{J_0}{ 3 r^3}+\Omega_{0}^{} + \alfo \omega_1 (r) + \dots ,  \label{rot1}\nonumber
\end{eqnarray}
where $\Omega_0$ is the integration constant that appears when solving for $\omega$ in (\ref{rotfull}) to the lowest order in $\alpha^{-1}$.
Inserting the above expansions in \eqref{rotfull} and using \eqref{nupeq} it is straightforward to read off an expression for $\om_1$; resulting in
\begin{eqnarray}
\om_1' &=& \frac{J_0}{r^4}\left( \la_1 + \nu_1 \right) + \frac{J_1}{r^4} \\
%=  \frac{J_0}{r^4}\int dr \left( \la_1' + \nu_1' \right) + \frac{J_1}{r^4} 
&=& - \frac{J_0}{r^4} \int dr \frac{(9 r_s-10 r) (r_s-2 r) \left(\frac{r \zeta_0 \Lambda}{r-r_s}\right)^{3/2}}{4 M_{p}^2 r (4 r-3r_s)^{5/2}}  + \frac{J_1}{r^4}, \nonumber
\end{eqnarray}
which can be integrated once to obtain $\om_1$ explicitly. For the discussion, it is sufficient the weak gravity limit result, given by
 \begin{equation}
%\om_1 =\frac{J_1}{3 r^3}-\frac{79  r_s   ({\zeta_0} \Lambda)^{3/2} J_0}{1728 M_{p}^2 r^{9/2}}-\frac{5  ({\zeta_0} \Lambda)^{3/2 } J_0}{56 M_{p}^2 r^{7/2}}-\frac{J_0 \nu _{10}^{}}{3 r^3}+\Omega _{10} ,\label{rotext}
\om_1 =\Omega _{10} +\frac{J_1-J_0 \nu _{10}}{3 r^3}-\frac{79({\zeta_0} \Lambda)^{3/2} J_0}{1728 M_{p}^2 r^{7/2}}
\left(\frac{8640}{4424}+\frac{r_s}{r}\right),\label{rotext}
\end{equation}
where $\Omega_{10}$ and $\nu_{10}$ are the integration constants for $\omega_1$ and $\nu_1$, respectively.  
It might appear that for small $r$ the corrections
associated to $\zeta_0$ become large, however we need to remember two things: i) this is valid only for $r>r_s$, 
ii)  $\zeta_0$ is proportional to $r_s$ in the weak gravity limit (see \eqref{zetadef}). In this limit, the term proportional to $\zeta_0$ is suppressed compared to the $J_0/r^3$ term from GR, by a factor $(r M_p)^{-1/2} (\alpha^{-2/3} \Lambda r_s)^{3/2}$. Thus the contribution from the cubic Galileon term to the rotation is highly suppressed. 

%%%%%%%%%%%%%%%%%%%%%%%%%%%%%%%%
\section{Coupling to matter}\label{couplingmatter}
%%%%%%%%%%%%%%%%%%%%%%%%%%%%%%%%
In the previous section we derived the vacuum solutions using the large $\alpha$ expansion. The integration constants in the solutions need to be determined by a coupling to matter. We include the matter coupling in the following way
\begin{equation}\label{coupledaction}
S_{lcg}=\int d^4 x \sqrt{-g}  \left( M_{p}^2 R + \frac{{1}}{2}g^{\mu\nu}\partial_\mu\Phi\partial_\nu\Phi+\frac{\al}{\Lambda^3}\Box\Phi g^{\mu\nu}\partial_\mu\Phi\partial_\nu\Phi \right) +2 S_{m}[\Psi_m; A^2(\Phi) g_{\mu\nu}].
\end{equation}
$S_M$ represents the action for some matter field $\Psi_m$ and $A^2(\Phi)$ is the conformal factor relating the metric in the Jordan and Einstein frames, $g^J_{\mu\nu} = A^2(\Phi) g_{\mu\nu}$. This conformal factor converts
the non-minimal coupling between $\Phi$ and the curvature $R$ into a $\Phi$-dependent modification of the
geodesic equations derived from \eqref{coupledaction}; this is usually referred as a \emph{fifth-force}.  
Actually for the simplest case of $2 \ln A = -\beta \Phi/M_p $, which corresponds to the Brans-Dicke theory when the Galileon term is absent, the Jordan frame version of \eqref{coupledaction} is given by 
\begin{align*}
S_{BDG} = &\int d^4 x \sqrt{-g_J } \left( \Phi_J M_p^2 R_J - \frac{M_p^2 \omega_{BD}}{\Phi_J} \partial^\mu \Phi_J\partial_\mu\Phi_J +  \frac{\al (2\omega_{BD}+3)^{3/2}}{4\sqrt\pi \Phi_J^3 \La^3}  \Box\Phi_J \partial^\mu \Phi_J\partial_\mu\Phi_J      \right)  \nonumber \\
&+2 S_{m}[\Psi_m; g_J{}_{\mu\nu}],
\end{align*}
where the pertinent redefinition of the scalar field and conformal transformation are 
\begin{equation*}
\frac{\partial_\mu \Phi_J}{\Phi_J}   =\sqrt \frac{1}{M_p^2(2\omega_{BD}+3)} \partial_\mu\Phi, \ \ \ g = \Phi^4_J  g_J ,
\end{equation*}
and $\beta^2 = (2\om_{BD}+3)^{-1}$. In the next section, we will also consider the case of $\ln A \sim \Phi^2/M_p^2$, which, for a normal scalar field, provides an interesting example in which the effective coupling between the scalar field and matter can be dramatically amplified in the interior of high density stars, due to the so-called scalarisation phenomenon \cite{Damour:1993hw}. We will focus on these two choices of $A(\Phi)$ and refer them as the linearly and quadratically coupled models, respectively. Moreover, we work exclusively in the Einstein frame. 

The equations of motion, derived from $\eqref{coupledaction}$, with respect to $\Phi$ and $g_{\mu\nu}$ are 
\begin{subeqnarray}\label{mattcoupledeqs}
\xi_{\mu\nu}&:=&M_{p}^2\left(R_{\mu\nu}-\frac{1}{2}g_{\mu\nu}R\right) 
 = \frac{1}{2} \gmn\covD^\alpha\Phi\covD_\alpha\Phi - \covD_\mu\Phi\covD_\nu\Phi\nonumber \\ & & -\frac{\al}{\Lambda^3}\left(\Box\Phi\partial_\mu\Phi\partial_\nu\Phi-2\partial_\mu\Phi\covD_\nu\covD_\alpha\Phi\covD^{\alpha}\Phi + \gmn\covD_\alpha\Phi\covD^{\alpha}\covD^{\beta}\Phi\covD_{\beta}\Phi\right) + \kappa T_{\mu\nu},\label{meqc}\\
\xi_\Phi&:=&-g^{\mu\nu}\covD_\mu\covD_\nu\Phi+2\frac{\al}{\Lambda^3}\left[\covD^\beta\covD^\nu\Phi \covD_\beta\covD_\nu\Phi-
(\Box\Phi)^2+ R^{\mu\nu}\covD_\mu\Phi\covD_\nu\Phi\right]=2 \gamma T. \label{sc1m}
\end{subeqnarray}

We have defined the energy momentum tensor $T^{\mu\nu}$ (with trace $T$) and the effective coupling strength between the scalar field and matter, $\gamma(\Phi)$, as 
\begin{eqnarray*}
\sqrt{-g}  T^{\mu\nu} &=&2  \frac{\delta S_m}{\delta g_{\mu\nu}},\\
 \gamma(\phi)&=&\frac{\partial \ln A}{\partial \Phi}.
\end{eqnarray*}
The quantity $\gamma$ traditionally plays a relevant role in constraining scalar-tensor theories within the Solar system using the parameterized post-Newtonian formalism (see for example \cite{Will:2014xja}). However, the first post-Newtonian parameters depend on the asymptotic behaviour of the scalar field, which is uknown in the context of the Galileon model, because we do
not know the exact exterior solution for the scalar field.  
The existence of an intermediate Vainshtein regime between the star and the asymptotically flat spacetime poses some difficulties in the
use of asymptotic values to constrain the model \eqref{coupledaction}.
%; this is because we do
%not know the exact exterior solution for the scalar field.   
In the next section, we solve the equations \eqref{mattcoupledeqs} for a matter source described as a perfect fluid. First, we consider a simple model 
characterized by a constant and low density, and then a
more realistic description of matter using a polytropic equation of state.  For the first case, it is sufficient to consider the linearly coupled model since, as shown in \cite{Damour:1995kt}, in
weak gravity any deviation from GR is sensitive only to the cosmological value of $\gamma$, which is precisely the field-independent coupling strength of the linear model. However, for the second case we consider the quadratically coupled
 model to allow for the presence of strong gravity deviations from GR, analogous to the ones reported in
\cite{Damour:1993hw, Damour:1996ke}. We focus on the first case next.
%%%%%%%%%%%%%%%%%%%%%%%%%%%%%%%
\subsection{Linearly coupled weak gravity model} \label{linandtov}
%%%%%%%%%%%%%%%%%%%%%%%%%%%%%%%
Let us consider an incompressible source of matter described by a perfect fluid with pressure $P(r)$ and constant density $\rho=\rho_0$. Its energy momentum tensor is given by  $T_{\mu\nu}=(\rho_0+P) u_\mu u_\nu +P g_{\mu\nu}$, where the 4-velocity $u_\mu$ of the fluid satisfies $u^\mu u_\mu = -1$. We consider the coupling function given by $\ln A^2(\Phi)={M_p^{-1}{\beta} \Phi}$. At the lowest order in the large $\alpha$ expansion, the equations \eqref{meqc} reduce exactly to those of GR, whose solution for the interior of a static and spherically symmetric distribution of matter with an exterior Schwarzschild spacetime is known as the Tolman-Oppenheimer-Volkoff (TOV) solution  
\begin{subequations}
\begin{align}
ds^2_{TOV}& = - e^{2\Psi} dt^2 + e^{2 \Xi} dr^2 + r^2(d\theta^2 + \sin^2\theta d\varphi^2), \label{lineelementtov} \\
e^{2\Xi} &= \frac{1}{1-2\frac{m(r)}{r}}, \nonumber \\
\frac{e^{\Psi}}{\nu_0} &= \frac{3}{2}\left( 1 - \frac{r_s}{R}  \right)^{\frac{1}{2}} - \frac{1}{2}\left( 1 - \frac{r_sr^2}{R^3}    \right)^{\frac{1}{2}}, \nonumber \\
m&=\frac{4\pi}{3 M_{p}^2} \rho_0 r^3, \nonumber \\
P_{TOV} &= \rho_0\left[  \frac{\sqrt{1-r_sr^2 R^{-3}} - \sqrt{1 - r_sR^{-1}}}{3\sqrt{1-r_s R^{-1}} - \sqrt{1-r_sr^2 R^{-3}}}      \right], \label{ptov}
\end{align}
\end{subequations}
where the radius of the star, $R$, is defined as the radius where the pressure $P_{TOV}$ vanishes,
and $\nu_0$ is an integration constant used to match the metric with Minkowski at $r=0$.
 Moreover, we have already used the matching at $R$ 
between the radial component of the Schwarzschild exterior solution, $\left( 1-\frac{r_s}{r} \right)^{-1}$,  and the radial component of the solution above. This matching is given by
\begin{equation}
r_s = \frac{8\pi}{3 M_{p}^2} \rho_0 R^3. \label{ltov}
\end{equation}

An incompressible equation of state is one of the few cases in which the 
equations of motion admit an analytic solution, and despite of its simplicity, it illustrates some general features of more realistic solutions. In particular, one finds a lower bound for the mass-radius ratio of the star by requiring that the central pressure remains finite, as can be seen from \eqref{ptov} by taking $r=0$. 

It is also possible to add a slow rotation to the TOV solution by adding the term
$2 r^2 \omega(r) \sin^2\theta d\varphi dt $ to the line element \eqref{lineelementtov}. We will do this in detail later, but we state in advance that the matching conditions determine the scalar charge and the angular momentum as follows:
\begin{subequations}\label{conservedquantitiesmatch}
\begin{align}
\zeta &   = \frac{1}{M_{p}^3}\int dr \sqrt{-g}\beta T +\frac{\sqrt{-g}|_0}{M_{p}^2\sqrt{-g}|_R} \left.\frac{\partial L}{\partial \Phi'} \right|_{r=0},\label{Qcon} \\
 J&=e^{-(\lambda(R) + \nu(R))}R^{4} \omega'(r)|_R  \label{Jcon}.
\end{align}
\end{subequations}
The first equation is obtained by integrating once the Euler-Lagrange equation for the scalar field inside the star and matching it with the scalar charge seen by the exterior solution. In the second equation, $J$ is the integration constant from the exterior solution as defined in \eqref{rotfull}.  Note that these equations for $J$ and $\zeta$ are valid at any order in $\alpha^{-1}$.

Up to now, the only effect of the Galileon interaction is via the scalar field suppression inside the Vainshtein radius, which allows 
us to recover the GR solution for the metric at the lowest order in $\alpha^{-1}$. In the next subsection, we show how the interior/exterior matching can be done consistently when the first $\alpha^{-1}$ corrections are included. 
%%%%%%%%%%%%%%%%%%%%%%%%%%%%%%%%%%%%%%%%%%%%%
\subsection{Corrections to the TOV solution} \label{galileontov}
%%%%%%%%%%%%%%%%%%%%%%%%%%%%%%%%%%%%%%%%%%%%%
Inserting the TOV solution in the scalar field equation of motion and using the expansion in terms of $\alpha^{-1}$, we can solve
the scalar field equation to the next order. In order to include the first order Galileon corrections, we assume that the pressure has the following form 
\begin{equation}
P=P_{TOV}+\alfo P_1, \label{pressuresplit}
\end{equation}
and the metric is given by  
\begin{equation}
g_{\mu\nu}^{int}dx^\mu dx^\nu = ds^2_{TOV} + \frac{1}{\alpha}ds^2_{G}= - e^{2\left(\Psi + \alfo \nu_1 \right)} dt^2 + e^{2\left( \Xi +\alfo \lambda_1 \right)} dr^2 + r^2(d\theta^2 + \sin^2\theta d\varphi^2). 
\label{tovcormet}
\end{equation}
In this section, we will consider a low density star and take the weak gravity limit. We assume that the presence of the scalar field does not modify the incompressible nature of the star, so $\rho=const.$. Expanding the 
scalar field equation of motion \eqref{sc1m} with $\Phi = \alpha^{-1} \Phi_1(r)$ and the metric \eqref{tovcormet} we can solve for $\Phi_1'$,
\begin{equation}
\Phi_{1}' =\pm\left( \frac{r}{2}\sqrt{\frac{\beta \La^3 \rho}{3 M_p}}+\frac{29 r^3 }{240}\sqrt{\frac{\beta\La^3\rho^3}{3 M_p^5}} \right). \label{sfintcor} 
\end{equation}
Just as for the exterior scalar field there are two branches of solutions. We pick the positive root since we want to match the solution with the positive branch of the scalar field outside the star to ensure that the asymptotic flat solution exists. Performing the same expansion in the metric equations \eqref{meqc}, we get a set of the equations for the metric
and pressure corrections, which yield 
\begin{align}\label{intsol}
\lambda_1  =&  \frac{r^4}{240}\sqrt{\frac{\beta^3 \La^3 \rho^3}{3 M_p^7}},    \nonumber \\
\nu_1=& -\frac{r^2(r^2-5R^2)}{160}\sqrt{\frac{\beta^3\La^3\rho^3}{3 M_p^7}}, \nonumber \\
P_1 = &-\frac{r^2-R^2}{8}\sqrt{\frac{\beta^3 \La^3 \rho^3}{3 M_p^3}}.
\end{align}
In \eqref{intsol} we have already fixed the integration constants for the interior solution by imposing regularity at $r=0$ and vanishing $P_1$ at $r=R$, i.e., we impose the star has the same radius as it would be without the Galileon interactions. This means that the contributions from the scalar field to the pressure must change the density. It is convenient to write down explicitly the exterior solution with all the integration constants that 
we need to fix at the surface of the star. As we are considering a low density star we can take the limit $r_s/r \ll 1$. The metric and scalar field outside the star are given by
\begin{align}
ds^2_{ext}& =g_{\mu\nu}^{ext}dx^\mu dx^\nu \nonumber \\
&=-\nu_{0}^{ext}\left(1-\frac{r_s}{r}\right)\left[1+\frac{2}{\alpha}\left(  -\frac{3 (\zeta_0\La)^{3/2}}{8 M_{p}^2 r^{1/2}}+\frac{\lambda _{10}^{{ext}}}{r}+\nu _{10}^{{ext}}  \right) \right] dt^2 \nonumber \\ &+\left(1 + \frac{r_s}{r}\right)\left[ 1+\frac{2}{\alpha} \left( -\frac{3  r_s (\zeta_0\La)^{3/2}}{128 M_{p}^2 r^{3/2}}+\frac{  (\zeta_0\La)^{3/2}}{16 M_{p}^2 r^{1/2}}-\frac{r_s \lambda _{10}^{{ext}}}{r^2}-\frac{\lambda _{10}^{{ext}}}{r}  \right) \right] dr^2 \nonumber \\
 & + r^2 d\theta^2 + r^2 \sin^2\theta d\varphi^2, \label{extmet} \\
\Phi_{1,ext}'  = &-\frac{7r_s}{16}\sqrt{\frac{\zeta_0\La^3}{r^3}}-\frac{1}{2}\sqrt{\frac{\zeta_0\La^3}{r}}.\label{extsf}
\end{align}
We use the label \emph{ext} to the integration constants of the exterior solutions  \eqref{phischw} and \eqref{weakcorrectionsexterior}. 
We note that once the Galileon interaction is turned on, there is  no reason for the density of the star to be the same as in the GR solution. Therefore, it is convenient to split the energy density as 
\begin{equation}
\rho = \rho_0 + \frac{1}{\alpha}\rho_1, \label{rhosplit}
\end{equation}
where $\rho_1$ is the contribution to the density coming from the Galileon interaction and $\rho_0$ is equal to the density of the pure GR solution. Also note that $\nu_{0}^{ext}$ is just a normalization factor related to $\nu_{0}$ in \eqref{lineelementtov}, which is already fixed so that the interior metric becomes Minkowski in the limit $r \to 0$, $\left.e^{2\Psi}\right|_{r=0}=1$. Therefore, we are left with four free parameters $r_s, \zeta_0, \lambda _{10}^{ext}$ and $\nu _{10}^{{ext}}$, which are determined by the matching conditions $\Phi_{1}'=\Phi_{1,ext}' , g_{rr}^{int}=g_{rr}^{ext}, g_{tt}^{int}=g_{tt}^{ext}$ and $g_{tt}^{int}{}'=g_{rtt}^{ext}{}'$,
at $r=R$. The results are 
\begin{align}
r_s=&\frac{R^3 \rho_0  }{3 M_{p}^2}+\frac{R^3   \rho_1 }{3 M_{p}^2\alpha},\nonumber \\
\zeta_0=&\frac{R^3 \beta  \rho_0}{3 M_{p}}, \nonumber \\
 \lambda _{10}^{ext}=&0,\nonumber \\ 
\nu _{10}^{{ext}}=&  \frac{R^4 \rho_0^{3/2}}{40}\sqrt{\frac{\beta^3\La^3}{3M_p^7}}, \nonumber \\
\nu_0^{ext}=&1+\frac{ R^2 \rho_0}{2M_p^2}+\frac{12M_{p}^2 R^2   \rho_1 +11 R^4 \rho_0   \rho_1}{24 M_{p}^4 \alpha}. \label{match1}
\end{align}
The correction to the mass due to the Galileon interaction is explicitly considered by the way in which we 
split the density $\rho$ and this is reflected in the expression for $r_s$ above, where we see that
the Schwarzschild radius is corrected (in pure GR, $r_s$ would be related only to $\rho_0$) even 
though $\la_{10}^{ext}=0$.  
The matching for $\zeta_0$, which is performed directly from the scalar field solutions \eqref{sfintcor} and \eqref{extsf} here, 
confirms the result \eqref{Qcon} and naturally extends the interpretation of
$\zeta_0$ given before in terms of a point mass living in a
Minkowski space-time. Indeed one can check that at the lowest order in $r_s/r$, the result \eqref{vr} for the Vainshtein radius in terms of the point mass is recovered. From \eqref{vr} and the matching for $r_s$ given above we can estimate
the condition for the TOV metric to be dominant over the corrections  \eqref{intsol}. In particular for the radial components we have
 \begin{equation}
\frac{r^8}{R^9} \ll \frac{r_V^3}{r_s^4}. \label{intmetcond}
 \end{equation}
The left hand side can be at most of order $1/R$ while the right hand side is much larger than $1/r_s$ as $r_V/r_s \gg 1$, thus \eqref{intmetcond} is easily fulfilled. 

This closes our study of low density static solutions. The Vainshtein mechanism works well inside the matter configuration and suppresses the scalar field effects in order to recover the GR solution with a negligible backreaction from the Galileon term, and this solution can be matched 
smoothly to the exterior metric - the Schwarzschild solution with small Galileon corrections. 

\subsection{Corrections to rotation}
Next we consider the solution for rotation in the weak field limit. Inside the star the equation of motion,
\begin{equation}
2 e^{2 \la} r (\rho +P) \om+M_{p}^2 \left(\left(-4+r \Xi'+r \Psi'\right) \om '-r \om ''\right)=0, \label{roteqint}
\end{equation}
has the solution
\begin{equation}
\omega=\Omega _{0}+\frac{ r^2 \rho \Omega _{0}}{5 M_p^2}, \label{introtgr}
\end{equation}
where one of the integration constants has been fixed to zero in order to obtain a regular solution at $r = 0$ and $\Omega _{0}$ represents a slowly rotating frame of reference with respect to the Minkowski spacetime at $r=0$ obtained by the coordinate transformation $t'=t, r'=r, \theta'=\theta, \varphi'=\varphi - \Omega_0 t$. The transition to the exterior slowly rotating solution, 
\begin{equation}
\omega_{ext}(r) = -\frac{J_0}{ 3 r^3}+\Omega_{0}^{ext},
\end{equation}
requires $\omega=\omega_{ext}$ and $\omega'=\omega_{ext}'$ at $r=R$, which in turn implies
\begin{equation}
\Omega _{0}^{ext} = \Omega _{0}+\frac{{R}^2   \rho  \Omega _{0}}{3 M_{p}^2},  \ \ \ J_0 =  \frac{2 {R}^5  \rho  \Omega _{0}}{5 M_{p}^2},
\end{equation}
for an arbitrary $\Omega_0\neq 0$. 

Now we consider the correction to this solution due to the Galileon term. Fro this, consider the metric 
\begin{align}
g_{\mu\nu}dx^\mu dx^\nu  =& - e^{2\left(\Psi + \alfo \nu_1 \right)} dt^2 + e^{2\left( \Xi +\alfo \lambda_1 \right)} dr^2\nonumber \\ & + r^2(d\theta^2 + \sin^2\theta d\varphi^2) +2 r^2 
\left( \omega(r)+\frac{1}{\alpha}\omega_1(r) \right) \sin^2\theta d\varphi dt , 
\label{tovcormet2}
\end{align}
together with the splitting \eqref{pressuresplit} and \eqref{rhosplit} for the density and pressure, respectively.
The first perturbative term for the rotation, $\alpha^{-1} \omega_1$, is completely determined by the corrections to the diagonal components of the metric and to the pressure \eqref{intsol}, and by requiring that the solution is regular at the centre $r\to 0$. As before, the scalar field contributes indirectly through $\la_1$ and $\nu_1$. Expanding \eqref{roteqint} up to 
$\mathcal O(\alpha^{-1})$ and solving for $\omega_1^{}$, we obtain 
\begin{equation}
\omega_1^{} = -\frac{r^2 \left(5 r^2-14 R^2\right)  \rho_0^{3/2} \Omega _{0,\text{}}}{560}\sqrt{\frac{ \beta^3  \Lambda ^3}{3 M_p^7}}.%-\frac{r^2 \left(10 r^4-39 r^2 R_s^2+42 R_s^4\right) \beta  \kappa  \sqrt{M_{p} \beta  \Lambda ^3} \chi ^5 \Omega _{0}^{}}{3360 \sqrt{3} M_{pl}^6}
\end{equation} 
On the other hand, the Galileon correction for the small rotation of the exterior solution is given by \eqref{rotext}, and 
its integration constants are to be fixed 
by requiring $g_{t\varphi}^{}=g_{t\varphi}^{ext}$, $(g_{t\varphi}^{}){}'=(g_{t\varphi}^{ext}{})'$ at the surface of the star, $r=R$.  
This is done straightforwardly and the result is
\begin{align}
%\Omega _{0}^{ext}=&\Omega _{0}^{}+\frac{R^2 \delta ^2 \kappa  \Omega _{0}^{}}{3 M_{p}^2},\nonumber \\
%J_0=&\frac{2 R^5 \delta ^2 \kappa  \Omega _{0,\text{}}}{5 M_{p}^2},\nonumber \\
\Omega _{10}=&\frac{R^4  \rho_0^{3/2} \Omega _{0}^{}}{48}\sqrt{\frac{ \beta^3  \Lambda ^3}{3 M_p^7}}+\frac{R^2   \rho_1  \Omega _{0,\text{}}}{3 M_{p}^2},\nonumber \\
J_1=&-\frac{R^7  \rho_0 ^{3/2}\Omega _{0}^{}}{70} \sqrt{\frac{ \beta^3  \Lambda ^3}{3 M_p^7}}-\frac{2 R^5   \rho_1 \Omega _{0}^{}}{5 M_{p}^2}.
 \end{align}
The condition to guarantee $\alpha^{-1}\omega_1\ll\omega_0$ is the same as  \eqref{intmetcond}.  Thus for a slow rotation, the Vainshtein mechanism operates successfully to suppress deviations from GR. 

%%%%%%%%%%%%%%%%%%%%%%%%%%%%%%%
\section{Quadratically coupled model}
%%%%%%%%%%%%%%%%%%%%%%%%%%%%%%%
In the previous section we studied a matter coupling characterized by  $\ln A^2(\Phi)={M_p^{-1}{\beta} \Phi}$, or equivalently by a constant coupling strength $2 \gamma=M_p^{-1} \beta$, which corresponds to a Brans-Dicke theory when the Galileon term is absent in the action.  Now we go to the next case, that of a quadratic coupling, leading to an effective coupling strength $\gamma\sim\beta \Phi/M_p^2$ with a dimensionless $\beta$. It has been shown in \cite{Damour:1993hw, Damour:1996ke} that this is 
the simplest case where a configuration
of matter, a neutron star in particular, can develop a scalar field large enough to produce significant deviations from GR in the strong gravity regime. This effect, called \emph{spontaneous scalarization}, is present independently from the rotation. For this reason, we limit ourselves to the study of static solutions to the equations \eqref{mattcoupledeqs} in the presence of a matter source described by a perfect fluid with a polytropic equation of state to model high density stars. 

%%%%%%%%%%%%%%%%%%%%%%%%%%%%%%%%%%%%%%%%%%%%%%%%%%%%%%%%%%%%%%%%%%%%%%%%%%%%%%%%%%%%%%%%%%%%%%
\subsection{Toy model for scalarization in standard Scalar-Tensor gravity}
%%%%%%%%%%%%%%%%%%%%%%%%%%%%%%%%%%%%%%%%%%%%%%%%%%%%%%%%%%%%%%%%%%%%%%%%%%%%%%%%%%%%%%%%%%%%%%
Here we follow \cite{Damour:1993hw} to motivate the idea that a $\Phi-$dependent effective coupling strength may lead to 
a peculiar behaviour of the scalar field inside a matter source.  We only deal with the scalar field equation of motion, \eqref{sc1m}
with $\alpha=0$. Basically we are studying a test scalar field on a 
flat space-time in the presence of a matter source whose coupling is determined by $\beta$ and the scalar field itself.  Then we need to solve
\begin{equation}
\Box\Phi = -4\pi \frac{\beta}{M_p^2} \Phi T,
\end{equation}
on a Minkowski background. Furthermore we assume  
\begin{equation}
-T = M \left( \frac{4}{3} \pi R^3  \right)^{-1} \equiv \frac{3 s M_p^2}{4 \pi R^2},
\end{equation}
where $R$ and $M$ are the radius and (ADM) mass of the star and the quantity $s$ defined by the last equality as $s=r_s/R$ is the self-gravity of the star. In the weak gravity limit one can expand the effective coupling strength schematically as $\gamma= \gamma_0(1+ a_1 s + a_2 s^2 )$,
where $a_i$ are finite parameters. This would suggest that even when the self gravity of a star is large, the effective coupling 
$\gamma$ remains small if the asymptotic $\gamma_0$ is small to pass the post-Newtonian constraints. However this result is obtained perturbatively in $s$ and the toy model 
summarized here shows that it does not necessarily hold in the strong gravity regime.

The scalar field profiles inside and outside a star are given by the solutions to
\begin{equation}
\frac{2}{r}\Phi' + \Phi'' = \begin{cases} 3\   \mbox{sign}(\beta)  |\beta|  s  {R}^{-2} \Phi & r<R , \\ 0 & r>R  \end{cases} \label{phicases}
\end{equation} 
The choice of sign for $\beta$  determines, a posteriori,  whether the
scalarization phenomena takes place ($\beta<0$) or not.  Assuming $\beta < 0$, and after imposing the regularity at $r=0$,  the solution for $r<R$ is given by
\begin{equation}
\Phi = \phi_c\frac{\sin \sqrt{3 |\beta| s {R}^{-2}} r}{\sqrt{3 |\beta| s {R}^{-2}} r} ,
\label{scalarisation}
\end{equation}
where the central value of the scalar field, $\phi_c$,  is determined  by matching both $\Phi$ and its derivative with the exterior solution obtained from \eqref{phicases}, $\Phi^{ext} = -r^{-1}{\phi_s} + \phi_0$, at the surface of the star. This gives a relation
between $\phi_c$ and the asymptotic value $\phi_0$, 
\begin{equation} 
\phi_c =  \frac{\phi_0}{\cos(\sqrt{3  |\beta| s})}\label{scala}. 
\end{equation}
For non-zero $\beta$ and $s$,  $|\phi_c| > | \phi_0 |$. In fact, if $\sqrt{3 |\beta| s} = \pi/2 $ the drastic amplification of the central scalar field happens 
and therefore the local coupling strength, $|\gamma| = |\beta| \phi_c/M_p^2$, is 
significantly large even if $\gamma_0$ in the perturbative expansion above
is vanishingly small. Evidently this a simplistic model and the amplification
may not be equally strong in a more realistic model, but the effect would still be present, enhancing deviations from GR in the strong gravity regime.

%%%%%%%%%%%%%%%%%%%%%%%%%%%%%%%%%%%%%%%%%%%%
\subsection{Scalarization - simplified Galileon model}\label{simple}
%%%%%%%%%%%%%%%%%%%%%%%%%%%%%%%%%%%%%%%%%%%%
Now, we explore the same toy model as before, but taking into account the Galileon term in the equation of motion for $\Phi$, namely
\begin{equation}
{} \Box\Phi+\frac{2\al}{\Lambda^3}\left[ -g^{\alpha\beta}g^{\mu\nu}\covD_\alpha\covD_\mu\Phi \covD_\beta\covD_\nu\Phi+ (\Box\Phi)^2-  R^{\mu\nu}\partial_\mu\Phi\partial_\nu\Phi\right]= -4\pi \frac{\beta}{M_p^2} \Phi T . \label{scalalpha0}
\end{equation}
For the  Minkowski metric  the $R^{\mu\nu}$ term does not contribute. 
If we employ our $\alpha^{-1}$ expansion to solve this equation, we find
\begin{equation}
\Phi =\phi_{c}+ \alfo\Phi_1(r) + \frac{1}{\al}\Phi_2(r) ..., 
\end{equation} 
with $\phi_{c}=const.$. At order $\alpha^0$, we are left only with the second and third terms while $\Box \Phi$ is pushed to $\mathcal O(\alpha^{-1})$, effectively suppressing the canonical kinetic part of $\Phi$. The solutions for $\Phi_1$ and $\Phi_2$ will give rise to some additive integrations constants. However, it is important that $\phi_{c}\neq 0$, to keep a contribution from $T$ at the lowest order in $\alpha^{-1}$, since such a term is necessary to get a regular solution as $r\to 0$. Moreover, the constant $\phi_c$ gives the coupling strength $\gamma = \beta \phi_c /M_p^2$ as before. Explicitly, the solutions obtained are as follows:
\begin{itemize}
\item For the interior region, $r< R$, equation \eqref{scalalpha0} gives
\begin{align}
&-\frac{4 \left({\Phi _1}'{}{}^2+2 r {\Phi _1}'{} {\Phi _1}''{}\right)}{r^2 \Lambda ^3}-\frac{\tilde\beta  \phi_{c} {}}{{R^2}}\nonumber \\ 
&+\alfo\left[{-\frac{\tilde\beta  \Phi _1{}}{{R^2}}-\frac{2 {} {\Phi _1}'{}}{r}-\frac{8 {\Phi _1}'{} {\Phi _2}'{}}{r^2 \Lambda ^3}-{} {\Phi _1}''{}-\frac{8 {\Phi _2}'{} {\Phi _1}''{}}{r \Lambda ^3}-\frac{8 {\Phi _1}'{} {\Phi _2}''{}}{r \Lambda ^3}}\right]=0,\label{eq95}
\end{align}
where $ \tilde\beta = -\mbox{sign}(\beta)  3|\beta | {s} $ is a dimensionless $\mathcal O(1)$ parameter.
When solving for $\Phi_1'$ and $\Phi_2'$, and for their respective integration constants by imposing the regularity of the scalar field at $r=0$, we find
\begin{subeqnarray}
\Phi_1' &=& \pm \sqrt{ \frac{-\phi_{c} {\tilde \beta} \La^3}{12 R^2}}  r,\\
\Phi_2' & =& {-\frac{1}{8} r  \Lambda ^3-\frac{r^3 {\tilde \beta}  \Lambda ^3}{80 R^2}}.\label{phiintnodiv}
 \end{subeqnarray}
The first order solution $\Phi_1'$ is similar to the first term of $\Phi_{1}'$
in the linearly coupled model \eqref{sfintcor}. This behaviour suggests that the scalarization phenomena does not occur in this model. The second order solution $\Phi_2'$ is independent of the choice of sign for $\Phi_1'$. The total scalar field inside the star can be written as 
\begin{equation}
\Phi = \phi_{c} \pm \alfo  \sqrt{ \frac{-\phi_{c} {\tilde \beta} \La^3}{12 R^2}}  \frac{r^2}{2}+\frac{1}{\al}\left( {-\frac{1}{16} r^2  \Lambda ^3-\frac{r^4 {\tilde \beta}  \Lambda ^3}{320 R^2}}\right),
\label{tmscalarfield}
\end{equation}
where the integration constants coming from integrating \eqref{phiintnodiv} have been set to zero.  This is our interior solution.
\item For $ r > R$ we are simply in vacuum and \eqref{scalalpha0} is a quadratic equation that can be solved for $\Phi'$ without the $\alpha^{-1}$ expansion, leading to
\begin{equation}
\Phi' = \frac{-r^2 \Lambda ^3\pm\sqrt{r} \sqrt{16 \al \zeta_0 \La^{3} +r^3  \Lambda ^6}}{8 \al r}.
\label{exact}
\end{equation} 
We take only the upper sign since it corresponds to the asymptotically flat solution. Now we further split this solution into the two regimes separated by the Vainshtein radius. 
\subitem{$i)$} $r \ll r_V$:
Here, and specially near the surface of the star, the solution is dominated by the Galileon contributions, which after using the $\alpha^{-1}$ expansion, reads
\begin{equation}
\Phi'_{r<r_V} \simeq\alfo\sqrt{\frac{\zeta_0 \La^{3}}{4 r }}-\frac{r  \Lambda ^3}{8 \al}.
\label{largeal}
\end{equation}
This is the solution in the Vainshtein regime. As we approach to $r_V$ the scalar field dynamics coming from the canonical kinetic term becomes relevant
and the solution must be connected to the asymptotically flat solution. However to see this within the $\alpha^{-1}$ series
it is necessary to take several more terms in the expansion.   
\subitem{$i)$} $r > r_V$:
Far away from the source and beyond the Vainshtein radius, we have the asymptotic behaviour typical of a standard Scalar-Tensor theory, given by
\begin{equation}
\Phi'_{r>r_V} \simeq\frac{\zeta_0}{r^2} \Rightarrow \Phi_{r>r_V}\simeq -\frac{\zeta_0}{r } + \phi_0.
\label{asymptotic}
\end{equation}
This is the exterior solution.
\end{itemize}

We want to relate $\phi_c$ in the interior solution, to the value of $\Phi$ at $R$ deep inside the Vainshtein radius and then to $\phi_0$ in the exterior solution. The last step cannot be done easily in general as our large $\alpha$ approximation is valid only for $r<r_V$ and the asymptotic solution is valid for $r>r_V$. Formally there is no overlaping region where the matching can be performed. For the particular case of Minkowski, $\phi_0$ can be related to $\phi_c$ using the exact solution. However once we consider the Schwarzschild exterior metric, this is no longer possible.   
Therefore, we will match the $\alpha^{-1}$ expanded solution 
(\ref{largeal}) with the asymptotic solution (\ref{asymptotic}) at $r=r_V$, 
and asses the inaccuracies caused by this matching using the exact 
solution.  

\begin{figure}
\includegraphics[width=0.45\textwidth]{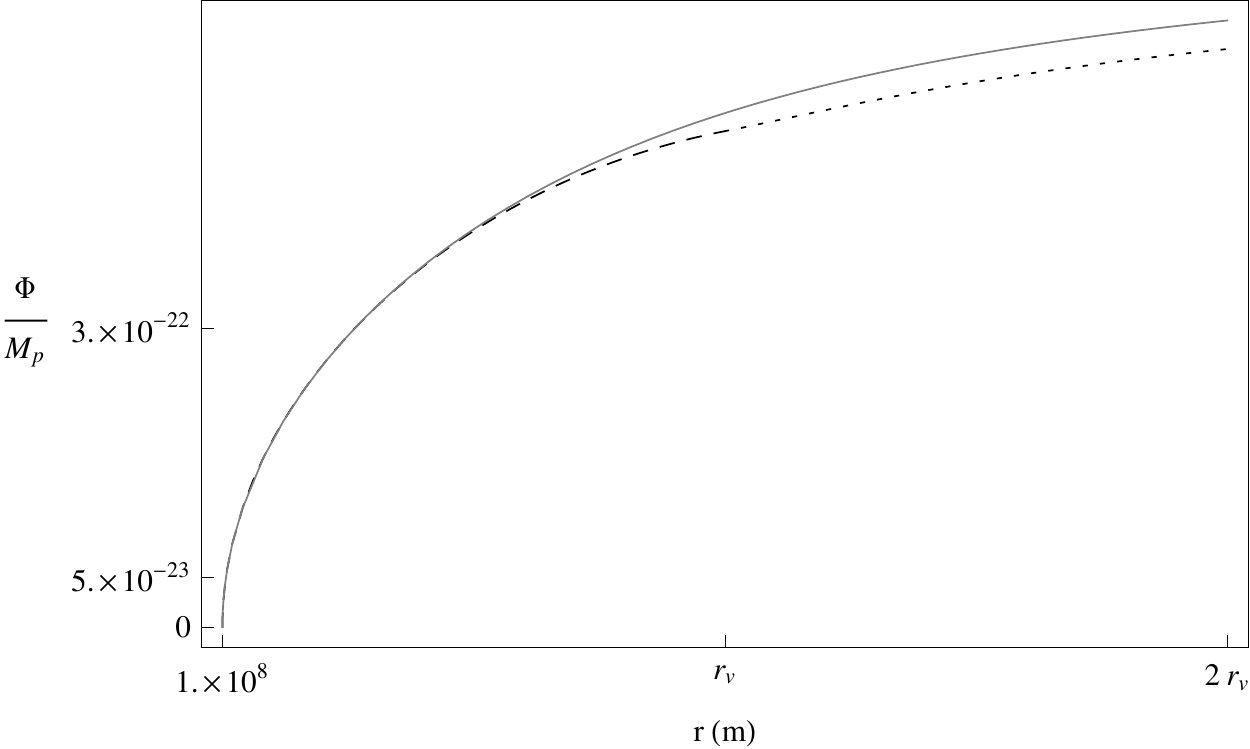}  \ \includegraphics[width=0.45\textwidth]{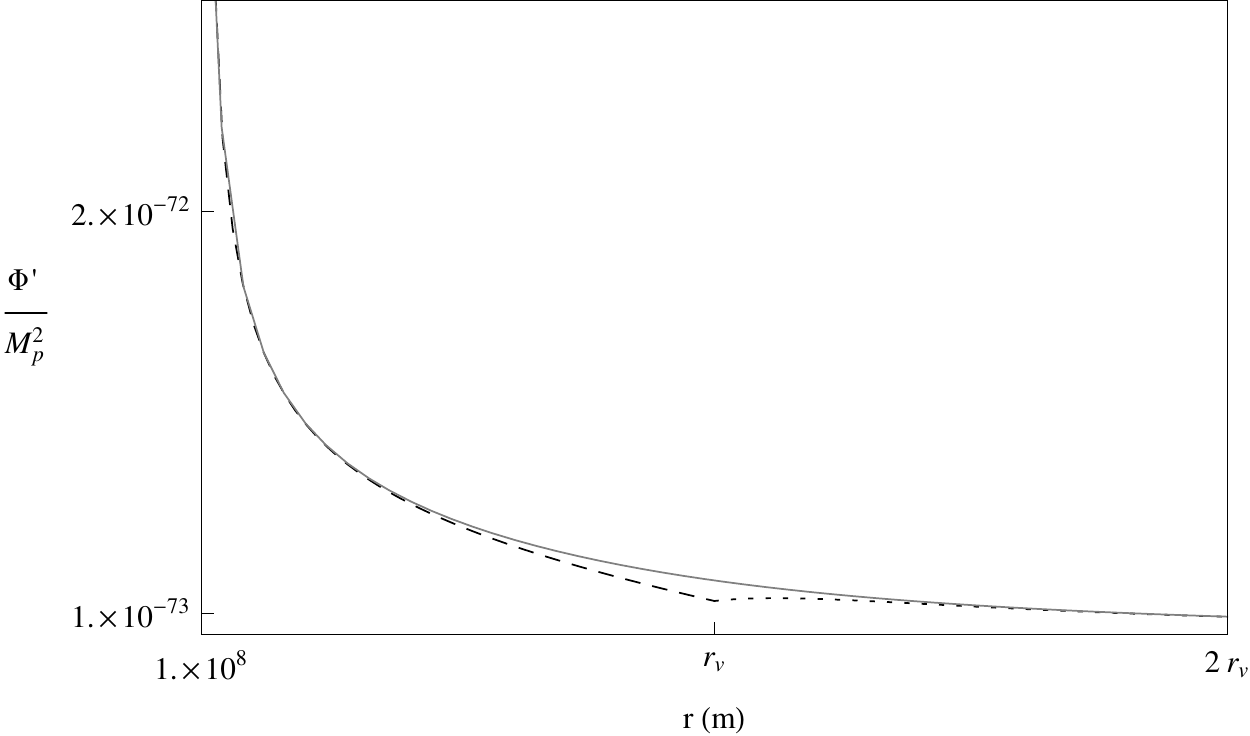}
\caption{ Exterior solutions for the scalar field and its derivative on a Minkowski space-time, with parameter fixed as:  $\La/\al^{1/3}=(1000\text{ Km})^{-1}$, $R=10^8$ m , $r_V = 10^{16}$ m . This corresponds to an object with the Schwarzschild radius $r_s  \sim 10^{-5} m$. The continuous line is the exact solution, while the dashed and dotted lines are respectively the $\alpha\to\infty$ approximation for $r<r_V$ and the $r\to \infty$ approximation for $r>r_V$.} 
\label{apprv}
\end{figure}

We impose the following matching conditions at $r=r_V$ and $r=R$ on the solutions $\Phi^I$, $\Phi^V$ and $\Phi^E$.  
\begin{subeqnarray}
\Phi^{\text V} |_{r_V}\simeq\Phi^{\text E}|_{r_V}
&\Rightarrow&\frac{\sqrt{\zeta_0\La^3 r_V}}{ \alpha}   -\frac{r_V^2  \Lambda ^3}{16 \al}+\phi_V =- \frac{\zeta_0}{r_V }+\phi_0, \\
\Phi^{\text V}{}' |_{r_V} \simeq\Phi^{\text E}{}' |_{r_V} &\Rightarrow&\alfo\sqrt{\frac{\zeta_0\La^3}{4 r_V }}-\frac{r_V  \Lambda ^3}{8 \al} = \frac{\zeta_0}{r_V^2},  \\
\Phi^{\text{I}} |_{R} \simeq \Phi^{\text{V}} |_{R} &\Rightarrow&\pm \alfo  \sqrt{ \frac{-\phi_{c} {\tilde \beta} \La^3}{12 R^2}}  \frac{r^2}{2}+\frac{1}{\al}\left( {-\frac{1}{16} r^2  \Lambda ^3-\frac{r^4 {\tilde \beta}  \Lambda ^3}{320 R^2}}\right) + \phi_c\nonumber \\
&& =  \frac{\sqrt{\zeta_0\La^3 {R}}}{ \alpha}   -\frac{{R}^2  \Lambda ^3}{16 \al} + \phi_V,\\ 
\Phi^{\text{I}}{}' |_{R}\simeq \Phi^{\text{V}}{}' |_{R} &\Rightarrow&\pm \alfo  \sqrt{ \frac{-\phi_{c} {\tilde \beta} \La^3}{12 R^2}}  {r}+\frac{1}{\al}\left( {-\frac{1}{8} r{}\Lambda ^3-\frac{r^3 {\tilde \beta}  \Lambda ^3}{80 R^2}}\right) \nonumber \\
&& = \alfo\sqrt{\frac{\zeta_0\La^3}{4 {R}}}-\frac{{R}{}\Lambda ^3}{8 \al}, \label{phicentralmatch}
\end{subeqnarray}
where the indices denote the region of validity of each solution: I for the interior, V for the Vainshtein regime and E for the exterior. 
From the first two equations we relate $\phi_0$ and $\phi_V$ at the Vainshtein radius,
\begin{equation}
\frac{2}{3}\left( \phi_0 - \phi_V \right) = \frac{\sqrt{\zeta_0\La^3 r_V}}{\alpha} - \frac{1}{8}\frac{r_V^2 {}\La^3}{ \al} = \frac{r_V^2\La^3{}}{8\al}; \label{phi0-phiv}
\end{equation}
and from the third and fourth equations we relate $\phi_V$ to $\phi_c$ at the radius of the star,
\begin{equation}
\phi_c = \frac{3}{4}\frac{\sqrt{\zeta_0\La^3 R}}{\alpha} -\frac{1}{\al}\frac{R^2 {\tilde \beta} \La^3}{320} + \phi_V.
\end{equation}
Putting these results together we have our final relationship between $\phi_c$ and $\phi_V$, that basically depends on the difference $\Phi^{V} |_{r_V} - \Phi^{V} |_{R}$ plus the following explicit contribution from the self-gravity of the star
\begin{equation}
{\phi_c -\phi_0}= \frac{\La^3}{ \al}\left[ \frac{3}{16}{} r_V^2\left( \frac{R^{1/2}}{r_V^{1/2}} -1  \right) + \frac{3 \beta R r_s  }{320} .    \right]\label{phicmink}
\end{equation}
Unlike the relation (\ref{scala}), the amplification of the central scalar field is not significant in this setup. Furthermore, the term proportional to $r_V^2$ is always the dominant one, thus the difference between the scalar field at the centre of the star and its asymptotic value does not depend explicitly on the properties of the star.  Since $\La^3/\al \sim \zeta_0 r_V^{-3}\sim M_p r_s r_V^{-3}$, the dimensionless quantity $M_p^{-1}(\phi_c-\phi_0)$ is of order $r_s/r_V \ll 1$, as can be confirmed from Fig.~\ref{apprv}. Therefore, the scalarisation does not happen in this case. 

As mentioned earlier, the matching at the Vainshtein radius is not well defined, but this is necessary since we cannot compute the solutions
for a general metric without the large $\al$ approximation. However for the Minkowski metric we do know the exact solution and therefore we can quantify the error due to this matching. If we perform the matching between the exact exterior solution given by \eqref{exact} and the approximate interior solution \eqref{tmscalarfield}, we find
\begin{equation}
\phi_c - \Phi(r\to\infty) = \int_{\infty}^R \Phi' dr + \frac{\Lambda^3}{\al}\left[\frac{R^2}{16}-\frac{1}{16}\sqrt{r_v^3 R + R^4} + \frac{3 \beta R r_s  }{320}   \right],
\end{equation}
where $\Phi'$ refers to the exact exterior solution \eqref{exact}. 

To get an idea of the error's magnitude, we consider the values $\La/\alpha^{2/3}=(1000\text{ Km})^{-1}$, $R=10^8$ m and $r_V = 10^{16}$ m, which fix $\zeta_0$
and also the Schwarzschild radius of the object through \eqref{zetadef}. The error in the relation between $\phi_c$ and $\phi_0$ is then given by 
\begin{equation}
\frac{\phi_c - \phi_0}{\phi_c - \Phi(r \to \infty)} -1 \simeq  0.4.
\end{equation}
This  relative error tells us that when we use the matching at $r_V$, we can trust the order of magnitudes of our results but not in the precise values. 
Fig.~\ref{apprv} shows a comparison between the approximated solution and exact solution. 

%%%%%%%%%%%%%%%%%%%%%%%%%%%%%%%%%%%%%%%%%%%%%%%%
\section{Compact Stars}
%%%%%%%%%%%%%%%%%%%%%%%%%%%%%%%%%%%%%%%%%%%%%%%%
In this section we solve the equations \eqref{mattcoupledeqs} under the $\alpha^{-1}$ expansion with a quadratic coupling for a stellar model consisting of a perfect fluid $T_{\mu\nu}=(\rho(r) +P(r)) u_\mu u_\nu +P(r) g_{\mu\nu}$ obeying a polytropic equation of state. The density and pressure are parametrised in terms of a dimensionless function $\chi(r)$, a polytropic exponent $\Gamma$, a polytropic constant $K$, and the formula
\begin{align}
\rho(r)& =  n_0 m_{b} \left( \chi(r) + \frac{K}{\Gamma-1} \chi(r)^\Gamma  \right), \\
p(r) & = K n_0 m_b \chi^\Gamma,
\end{align}
where $n_0= 1\times 10^{44} \ {m}^{-3}$ is the number density and $m_b=1.66 \times 10^{-21}$  kg is the baryon mass. From now on we stop using natural units, in order to give values of the densities in a way that is standard in the literature discussing compact stars in GR.
For a polytropic exponent $2\lesssim \Gamma\lesssim 3$ the observational mass-radius curves of neutron stars are well reproduced within GR. For smaller values of $\Gamma$ (around $ 5/3$), the solution corresponds to lower density stars, commonly defined by $c^2 n_0 m_b \chi < 5\times10^{14} \text{gr}/\text{cm}^3 \equiv\tilde \rho_0 $. For these low density stars, the value of $K$ can be estimated from the non-relativistic limit of a Fermi gas model, which yields a polytropic equation of state with $ \Gamma_{low}=5/3$ and a polytropic constant given by
\begin{equation}
\frac{K_{low}}{(n_0 m_b)^{2/3}} = \frac{(3\pi^2)^{2/3}}{5}\frac{\hbar}{m_b^{8/3}}.
\end{equation} 
Compact stars can be modelled in a more sophisticated way by considering them as having both a low density and a high density region, i.e. by taking a two-component polytrope. In this case the continuity of the pressure at the interface between these two components requires that for the high density region
\begin{equation}
\frac{K}{n_0^{\Gamma-1} m_b^{\Gamma-1}}=\frac{1}{c^{2\Gamma}}\frac{(3\pi^2)^{2/3}}{5}\frac{\hbar^2}{m_b^{8/3}} \rho_0^{\frac{5}{3} - \Gamma}.
\end{equation}
Although we consider a single component star of high density, the previous relation is useful to estimate the appropriate $K$ for a given $\Gamma$. We use $\Gamma=2.34$ because with this choice, the maximum mass of a neutron star, shown in the results below, is consistent with the maximum mass computed with a more realistic equation of state in agreement with observational data, e.g., EOS II in \cite{1985ApJ...291..308D}. % and with the observational data. 

%%%%%%%%%%%%%%%%%%%%%%%%%%%%%%%%%%%%%%%%%%%%%%%%%%%%%%%%%%%%%%%%%%%%%%
\subsection{Equations of motion}
%%%%%%%%%%%%%%%%%%%%%%%%%%%%%%%%%%%%%%%%%%%%%%%%%%%%%%%%%%%%%%%%%%%%%%
In order to easily compare with the analysis of \cite{Damour:1996ke}, it is convenient to write the metric explicitly in terms of a Schwarzschild-like radial component by redefining the $\mathcal O(\alpha^0)$ part of the metric as $e^{-2\la_0} =1-2\mu(r)/r$. Since the exterior metric solution asymptotes a flat spacetime, $\mu(r\to\infty)$ can be interpreted as the ADM mass. Up to $\mathcal O(\alpha^{-1})$, the metric is written as 
\begin{equation}\label{neumet}
ds^2=-e^{\nu(r)+\frac{2}{\alpha}\nu_1(r) }  dt^2 +
\frac{e^{\frac{2}{\alpha}\la_1(r) }}{1-\frac{2\mu(r)}{r}}dr^2 + r^2 d\Omega.
\end{equation}
We also redefine $\Phi$ to make it dimensionless, $\Phi \to M_{p} \Phi$, and write the pressure and 
density in terms of their Jordan frame counterparts, i.e., $(\rho, p)\to A(\Phi)^4(\rho, p)$ where $A(\Phi)=\exp(\beta \Phi^2/2)$. As before, we 
expand the scalar field as  $\Phi(r) = \phi_c + \alpha^{-1}\Phi_1(r)$. Inserting the above expansions and redefinitions in the equations of motion \eqref{meqc}, we got
\begin{subeqnarray}
\mu '(r)&=&\frac{4 e^{2\phi_c^2 \beta } G \pi  r^2 \rho(r)}{c^4}+\mathcal O(\alpha^{-1}), \\ 
{\nu}'(r)&=&\frac{8 e^{2\phi_c^2 \beta } G \pi  r^2 {p}(r)}{c^4 (r-2 \mu (r))}+\frac{2 \mu (r)}{r (r-2 \mu (r))} +\mathcal O(\alpha^{-1}),\\
{p}'(r)&=&-\frac{1}{2} \left({p}(r) +{\rho}(r) \right){\nu}'(r)+\mathcal O(\alpha^{-1}) . \label{neutrongalileon}
\end{subeqnarray}
The last equation can be obtained either as a combination of the Einstein equations or directly from the
conservation of the energy momentum tensor.
All the dynamical effects of the scalar field are removed, and its only contribution is a constant rescaling of 
the density at the lowest order. As we saw in the toy model of the previous section, $\phi_c$ is required to a obtain regular 
solution for the scalar field as $r\to 0$, which introduces an effective Newton's constant given by $e^{2 \phi_c^2 \beta} G$. 

The first order correction to the scalar field $\Phi_1$ is determined again using \eqref{sc1m}, and results in
\begin{eqnarray}
(r-2\mu)\Phi_1'\Phi_1''&= &\frac{e^{2 \phi_c^2 \beta } G \phi_c \pi  r^3 \beta  \Lambda ^3 (3 {p}{}-{\rho}{})}{{} \left(4 e^{2 \phi_c^2 \beta } G \pi  r^3 {p}{}+c^4 (2 r-3 \mu {})\right) } \nonumber \\
 & +&{\Phi_1'{}^2}{} \left(-1-\frac{2 e^{2 \phi_c^2 \beta } G \pi  r^2 ({p}{}-3 {\rho}{})}{c^4}-\frac{c^4 (r-2 \mu {})^2 \left(-4+r^2 \nu''{}\right)}{16 e^{2 \phi_c^2 \beta } G \pi  r^4 {p}{}+c^4 r (2 r-3 \mu {})}\right). \label{sfeqns}
\end{eqnarray} 
Just like in the Minkowski case, the effective coupling strength between the scalar field and matter is constant, and given by $\phi_c \beta$. We approximate the initial conditions for $\mu$ and $\nu$ at $r=r_{min}\approx 0$ 
by taking the functions on the r.h.s. of \eqref{neutrongalileon} as constants over an infinitesimal interval
$r_{min}+\delta r$, and then we integrate with respect to r to obtain
\begin{eqnarray*}
\mu_c& =& \frac{4 G {m_b} {n_0} \pi  r_{min}^3 \left({\chi_c}+\frac{{K} {\chi_c}^{\Gamma }}{-1+\Gamma }\right)}{3c^4} ,\\ 
\nu_c & = & \ln\left[1-\frac{2\mu_{c}}{{r_{min}}}\right]+\frac{4 e^{2 \phi_c^2 \beta } G K m_b n_0 \pi  \left(r_{min}^2+4\mu_{c} ({r_{min}}+2 \ln[{r_{min}}-2\mu_{c}]\mu_{c})\right) \chi ({r_{min}})^{\Gamma }}{c^4},
\end{eqnarray*}
where all the subscripts $c$ stand for quantities evaluated at $r_{min}$.
The same procedure for the scalar field requires that $\Phi_1'(r=0)=0$. However, in order to avoid numerical singularities we have to set $\Phi_1'(r_{min})\sim 0$, and in order to keep consistency with the large $\alpha$ approximation, we also 
need $\Phi_1(r_{min})\ll \phi_c$. 
We vary the initial condition for the density, $\chi_c$, in the range from $0.1$ to $50$, in units of $c^2/10 m_b n_0 \tilde\rho_0$. 
%to $\sim 50 c^2/ m_b n_0 \tilde\rho_0$. 
These choices are of order one below and above the typical density for a neutron star. Notice that the initial condition $\phi_c$ can be different for each density, however and for a preliminary analysis, we keep it the same for all the densities, with an arbitrarily chosen value of $\phi_c \sim 0.05$. Later, we will take into account the variation of $\phi_c$
in order to match a given boundary condition (for example an asymptotic value of the scalar field at infinity). In addition, the possible variation of the initial condition for $\Phi_1$ is neglected since it can be absorbed in $\phi_c$. Furthermore, to fix a value for the Galileon coupling constant we make use of the weak gravity estimate of $\Lambda^3/\al=M_{p} H_0^2$, where
$H_0$ is the current Hubble parameter. This estimate comes from
the requirement that the Galileon modifications are relevant for the present day expansion of the Universe, and at the same time they are consistent with constraints on GR deviations within the Solar System \cite{Nicolis:2008in}. This means $\Lambda^3/\al\sim (1000\  \text{Km})^{-3} $, giving a Vainshtein radius of the order of $10^{18}$m for the Sun, as can be computed from \eqref{rv}. For comparison, the aphelion of Neptune's orbit is of order $10^{12}$m, therefore the Solar System is well inside the region where the nonlinear Galileon dynamics takes place. From now on, we set $\alpha=1$ and by doing this we are treating $\Lambda$ as a meaningful physical parameter. Note that, as shown after \eqref{appdual}, our approximation is controlled
by the ratio $r/r_V$ and not $\alpha$ or $\Lambda$ themselves. 

\begin{figure}
\includegraphics[width=0.7\textwidth]{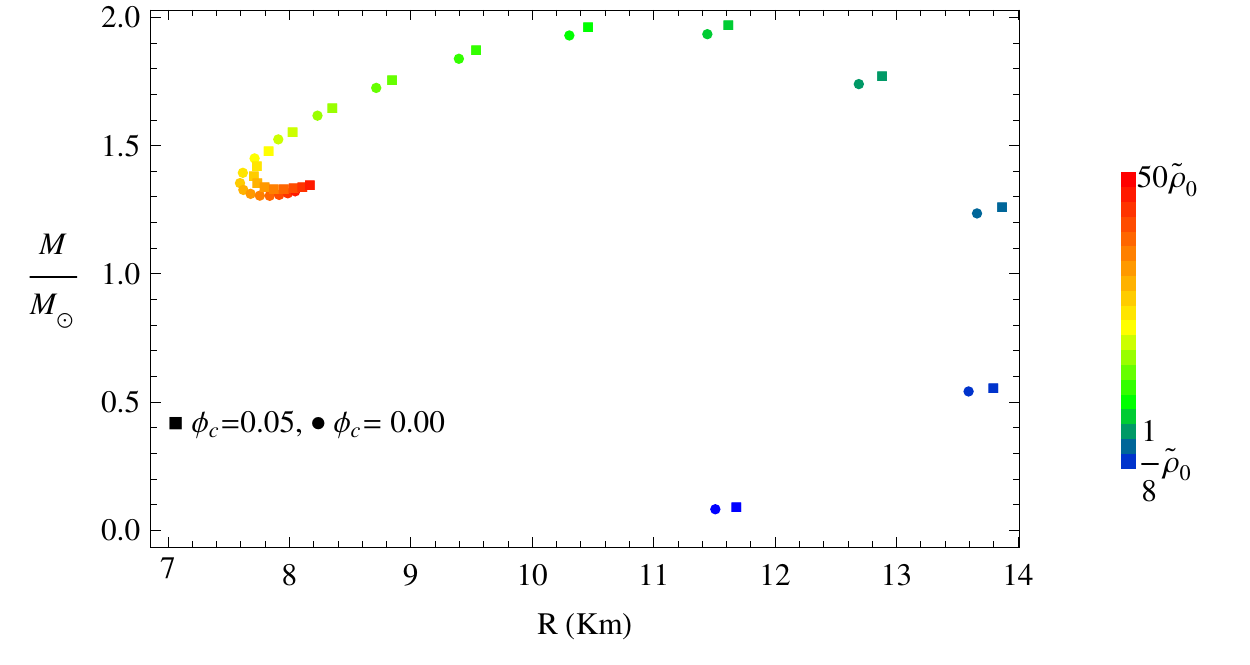}\caption{Mass-Radius curves for $\phi_c=0$ (GR) and 
$\phi_c=0.05$, $M_\odot$ stands for one solar mass. Each dot or square represents an equilibrium solution to the equations \eqref{neutrongalileon}, without the $\alpha^{-1}$ terms. The initial conditions for the density and pressure of each configuration are dictated by $\chi(r_{min})$ which changes from  $c^{-2}n_0^{-1}m_b^{-1} \frac{\tilde\rho_0}{8}$ to  $c^{-2}n_0^{-1}m_b^{-1} 50\tilde\rho_0$}\label{mrcurve}
\end{figure}

\begin{figure}
\includegraphics[width=0.7\textwidth]{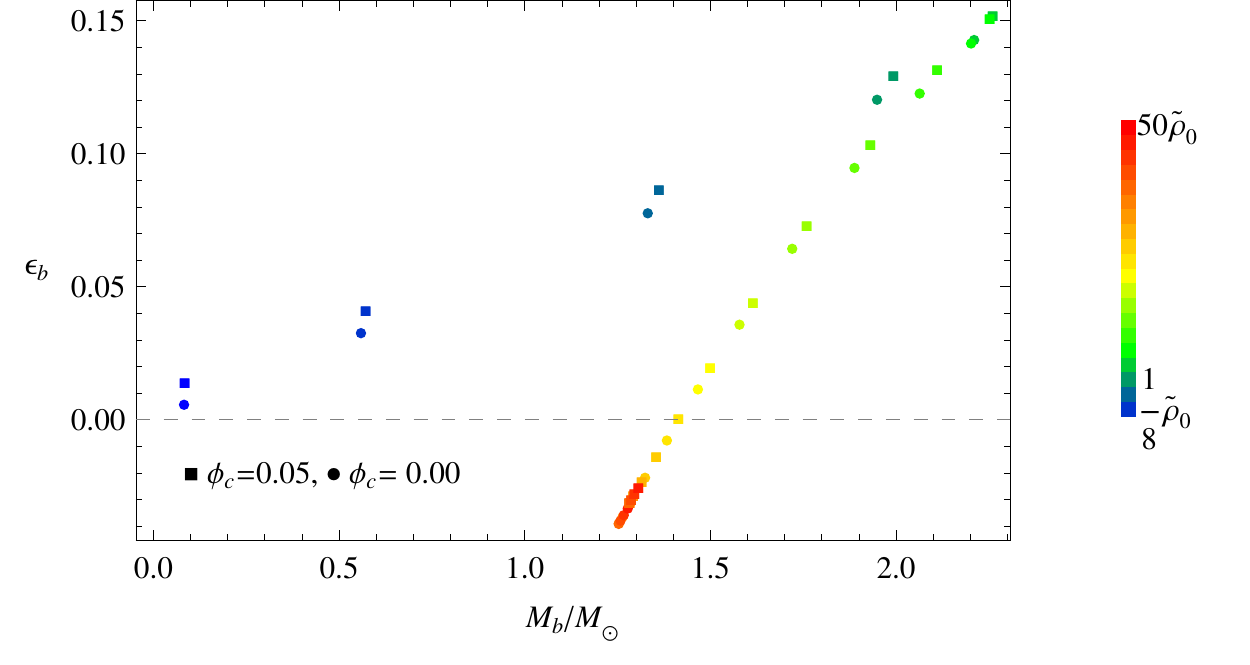}\caption{Fractional binding energy for the same set of solutions displayed in Fig. \ref{mrcurve}. }\label{becurve}
\end{figure}

We organize our approach in three steps; first we look for static configurations in the presence of the constant term $\phi_c$ in the scalar field solution. Then we study the dynamical part of the scalar field, $\Phi_1$, and finally we check the back-reaction of $\Phi_1$ onto metric. To characterise the static solutions, we use two quantities that are standard in the study of neutron stars: the fractional binding energy, $\epsilon_b$, and the baryonic mass, $M_b$. The fractional binding energy is defined by
\begin{equation}
\epsilon_b = \frac{M_b-M}{M},
\end{equation}
and it measures the difference between the gravitational mass $M$ and the mass that the dispersed baryons of the star would have, i.e. the total number of baryons in the star multiplied by $m_b$. With our 
particular choice of the quadratic coupling (which comes into play through the above redefinition of the density), $M_b$ is given by
\begin{equation}
M_b = m_b\int_0^R 4\pi n_0 e^{\frac{3\beta}{2}\Phi^2} r^2 \chi(r)\left(1-\frac{2\mu(r)}{r}\right)^{-1/2} dr.
\end{equation} 
A necessary condition for a star to be stable against radial perturbations is that its mass must increase as the central density increases, $d M/d \rho_c>0$. This is known as the \emph{static stability criterion} \cite{haensel2007neutron}. The stability changes at the turning point of the mass-radius curve, as can be seen by comparing Fig.~\ref{mrcurve} and Fig.~\ref{becurve}; from low to high density, all the configurations before the turning point are stable (i.e. the fractional binding energy becomes larger as the density increases), while the ones after the turning point are unstable. The maximum mass corresponds to the turning point of the binding energy curve. From the same figures we also see that a large $\phi_c$ would have a dramatic effect on the observable characteristic of a neutron star.  This means that the leading order scalar field coupling $\beta \phi_c$ must be small. In the next section we study the behaviour of the first order correction to the scalar field solution $\Phi_1$. 

\begin{figure}
\includegraphics[width=0.7\textwidth]{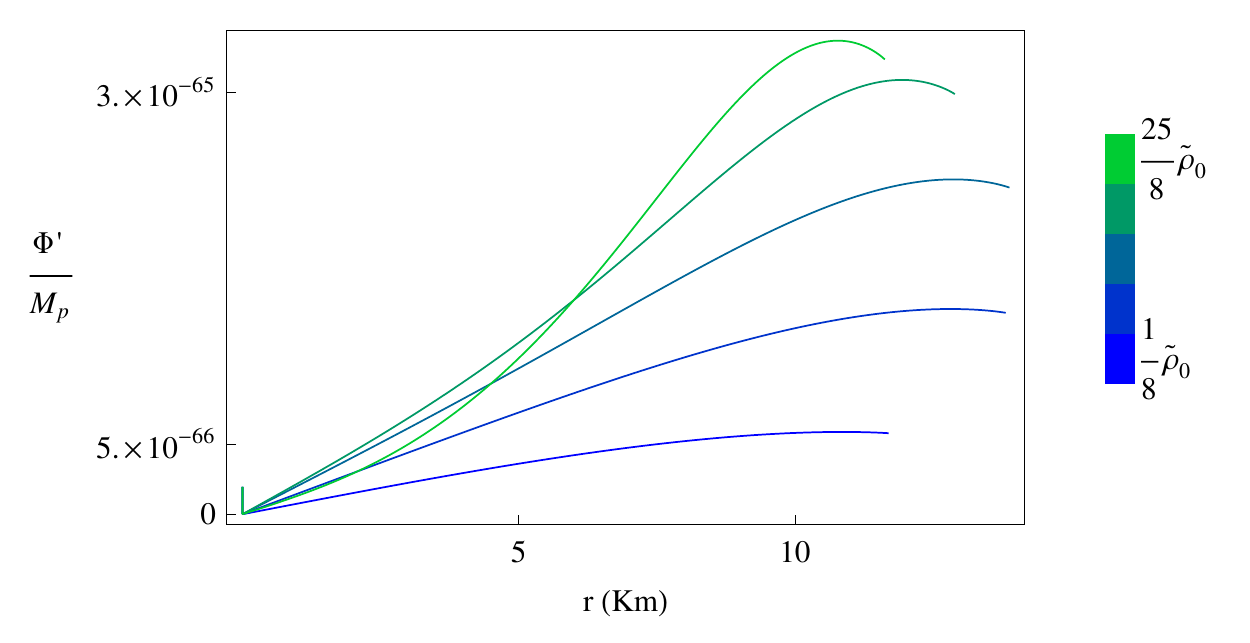}\caption{
Profiles for $\Phi_1'$ inside the five lowest density configurations of Fig.~\ref{mrcurve} and Fig.~\ref{becurve} for $\phi_c=0.05$, the next density in the sequence of solutions does not admit a scalar field solution. }\label{phipunstable}
\end{figure}
%%%%%%%%%%%%%%%%%%%%%%%%%%%%%%%%%%%%%%%%%%%%%%
\subsection{The dynamics of the scalar field inside the star}
%%%%%%%%%%%%%%%%%%%%%%%%%%%%%%%%%%%%%%%%%%%%%%
Let us focus on the configurations before the turning point of the binding energy curve (i.e. those with mass lower than the maximum mass) so that we have stable solutions. In Fig.~\ref{phipunstable} we show the solution for the scalar field for these stable configurations.  As we mentioned before, $\phi_c$ should be seen as an initial condition which can be different for every density, thus in principle, it is possible to engineer a set of solutions with small deviations from GR for low densities (i.e. a small $\phi_c$) and noticeable deviations for high densities. However, as we show below, under some physical requirements 
such a behaviour cannot be achieved. 
We assume the following:
\begin{itemize}
\item There exists an asymptotically flat solution outside the Vainshtein radius. This is supported by the results of 
\cite{Kaloper:2011qc}. In such a region, typically on cosmological scales, a bound on the asymptotic value of the scalar field, $\phi_0$, can be imposed. 
Note that $\beta \phi_0$ measure the coupling strength between the scalar field and matter in this region. 
\item The solution for the lowest density considered in our analysis saturates that bound.
\end{itemize}
We ask how large are the deviations from GR, i.e. what is the value of $\phi_c$, when we require to keep the same $\phi_0$ as we move from lower to higher density solutions. This is basically the same set up that leads 
to the discovery of strong gravity effects in the Brans-Dicke gravity \cite{Damour:1993hw}. The difference is 
that in order to evaluate the asymptotic scalar field $\phi_0$ we need to consider the intermediate Vainshtein regime between 
the star and the asymptotically flat region. This also implies that the constraint on $\phi_0$ does not come from the Solar System test in our case unlike in Brans-Dicke gravity. In order to compute $\phi_0$ we need explicit expressions for the matchings of the scalar field solution 
both at $R$ and $r_V$. Beyond the Vainshtein radius, the usual scalar tensor theory solution, without Galileon, can be used, hence for $r>r_V$, we assume the general static, spherically symmetric vacuum solution is
\begin{align}
ds^2 &= -e^{\nu_{ext}} dt^2 + e^{-\nu_{ext}}\left[ d\tilde r^2 + (\tilde r^2- a\tilde r)(d\theta^2 +\sin^2\theta d\varphi^2)  \right], \\
e^{\nu_{ext}} &= \left( 1-\frac{a}{\tilde r}\right)^{\frac{b}{a}}, \nonumber\\
\Phi_{ext}(\tilde r)& = \phi_0 + \frac{d}{a}\ln \left(1-\frac{a}{\tilde r} \right), \nonumber
\end{align}
where $a, b$ and $d$ are integration constants subjected to the condition $a^2-b^2=4d^2$. Note that this 
solution is in a gauge $g_{\tilde r \tilde r} = g_{tt}^{-1}$, whereas the solutions for the star and 
the Vainshtein region are
in the gauge $g_{\theta\theta} = r^2$. To match quantities computed in those different gauges we need the relation between $r$ and $\tilde r$, given by
\begin{equation}
r = \tilde r \left( 1-\frac{a}{\tilde r}  \right)^{\frac{a-b}{2a}}.
\end{equation} 
For the scalar field in the Vainshtein region, we use equation \eqref{phischw}, with the positive sign to ensure that it is compatible with 
an asymptotically flat solution. The procedure is summarised as follows: we solve numerically for the scalar field
inside the star, and then we match the solution at $r=R$ with the scalar field solution in the Vainshtein regime. We compute the Vainshtein radius using \eqref{vrs} and by mathcing the solution at $r=r_V$ to the exterior solution, we obtain $\phi_0$.  After doing this for the lowest density solution, we require that all the solutions for higher densities have the same $\phi_0$. This is achieved by changing the initial condition $\phi_c$ at every different density. With this procedure, we obtain
\begin{equation}
\phi_0 = \phi_c + \alfo \Phi^I_1|_{R} +  \Phi|_{r_V} -\Phi|_{R} +\left[\frac{2\Phi''}{\sqrt{(\nu')^2+4(\Phi')^2}}\left. \arctanh \frac{\sqrt{(\nu')^2+4(\Phi')^2}}{\nu'+2 r ^{-1}} \right]\right|_{r_V},
\end{equation}
where $\Phi^I_1|_R$ is evaluated using the numerical solution for the interior of the star, and all the other $\Phi$ refer
to the scalar field solution in the Vainshtein region. 

\begin{figure}
\includegraphics[width=0.7\textwidth]{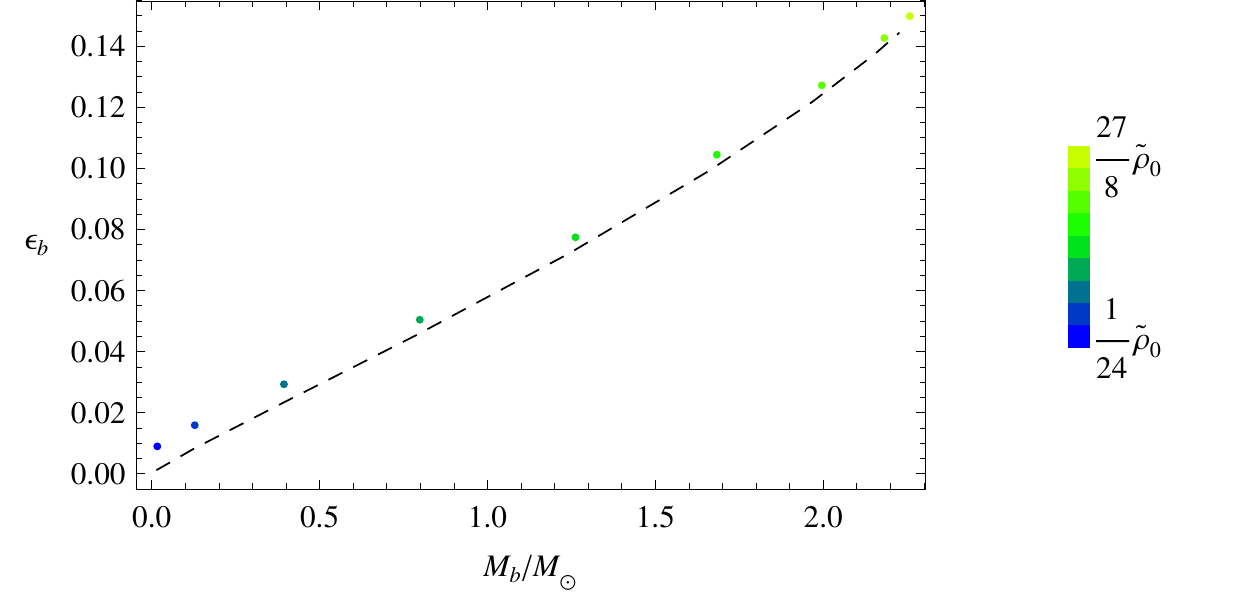}\caption{This plot shows the fractional binding energy curves for GR (dashed line) and for the cubic Galileon. In this plot we do not use realistic values for $\phi_c$ and $\La$. $\phi_c=0.050$ is used for the lowest density solution in order to show a 
significant deviation from GR and the value of $\La$ has been increased in order to amplify
the effects of the Galileon. We demand all the solutions to have the given $\phi_0\approx 0.050$, computed for the lowest density solution. The scalar field profiles that fulfil this condition require a smaller value of $\phi_c$ as the density increases. Hence the high density solutions are closer to GR solutions. }\label{fig:togr}
\end{figure}

\begin{figure}
\includegraphics[width=0.45\textwidth]{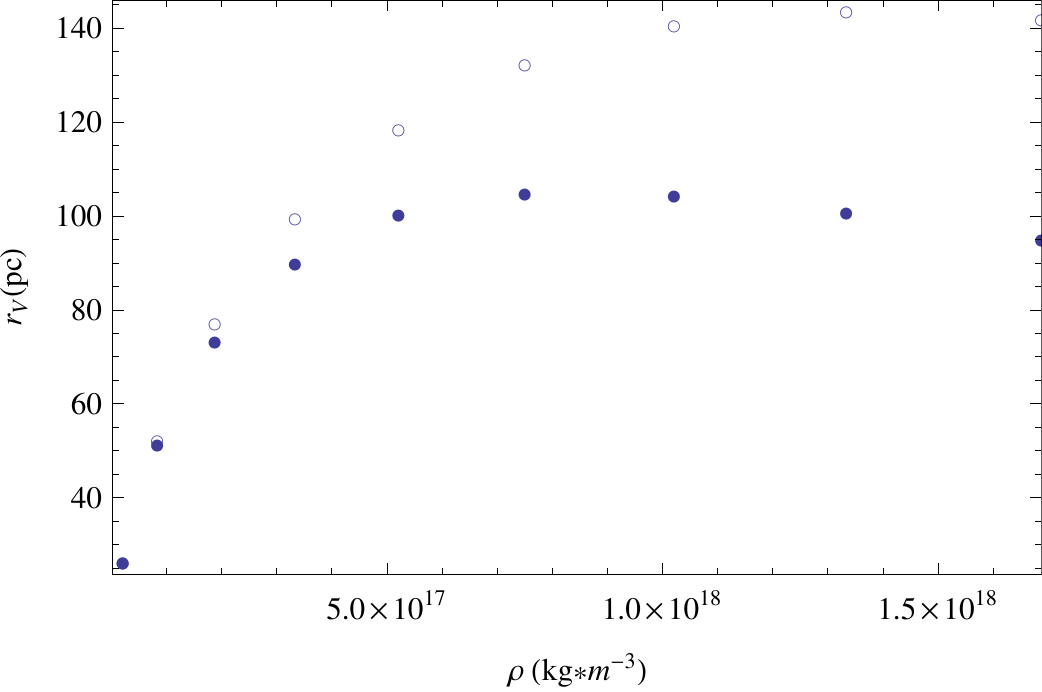} \includegraphics[width=0.45\textwidth]{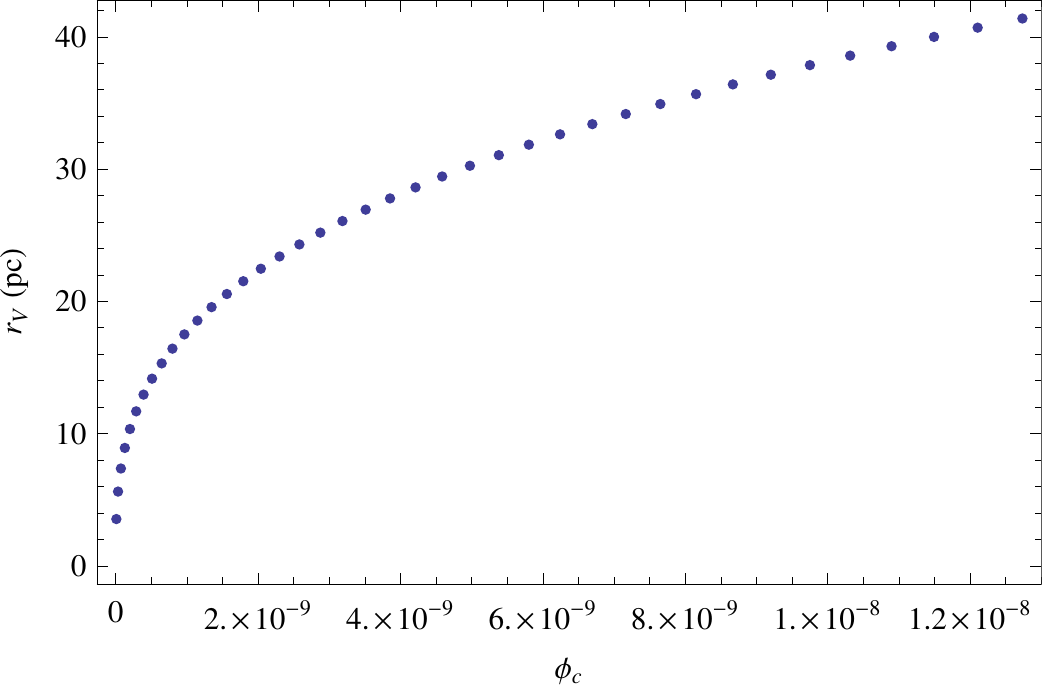}\caption{
Dependence of the Vainshtein radius on the density and $\phi_c$.  
Left panel: $r_V^{weak}$ (empty circles) and $r_V$ as a function of the central density, with $\phi_c$ fixed. $r_V^{weak}$ denotes the Vainshtein radius computed with the weak gravity approximation for the exterior scalar field.   
Right panel: $r_V$ as a function of $\phi_c$, for a fixed density. }\label{rvmod}
\end{figure}

The results are shown in Fig.~\ref{fig:togr}, where large 
values of $\phi_c$ and $\La$ were chosen so that the 
small effects for higher density solutions are visible. A detailed understanding of these results can be achieved with Fig.~\ref{rvmod}. 
Since $\Phi$ in the Vainshtein region grows as $\sqrt{r}$ in weak gravity limit, then the dominant term in $\phi_0$ is 
$\Phi|_{r_V} \sim M_p^{-1} \sqrt{\La^3 \zeta_0 r_V} \sim M_p^{-1} \La^3 r_V^2$. In Fig.~\ref{rvmod}, we see that at low densities $r_V$ grows monotonically with both, the density and $\phi_c$, so in a rough approximation $\phi_0\propto M_p^{-4}\rho \phi_c$. For this reason, if we want 
to keep $\phi_0$ constant as we increase the density, $\phi_c$ has 
to decrease, pushing the solutions towards the GR limit (remember $\phi_c$ measures the GR deviations inside the star). At higher densities, the dependence of $r_V$ on $\rho$ changes drastically due to strong gravity effects, and the Vainshtein radius begins to decrease as the density grows. However this is not enough to trigger large deviations from GR in our results.

At this point we have learned that in the cubic Galileon model a neutron star can carry a scalar field and still be indistinguishable from a neutron star in GR. The corrections to the metric of order $\alpha^{-1}$ are expected to be small too. In the next section we will confirm that this is the case.

%%%%%%%%%%%%%%%%%%%%%%%%%%%%%%%%%%%%%%%%%%%%%%%
\subsection{Corrections to the metric}
%%%%%%%%%%%%%%%%%%%%%%%%%%%%%%%%%%%%%%%%%%%%%%%
So far, all the computations concerning the Einstein equations have been done taking only the leading order contributions in the large-$\alpha$ approximation. Now we would like to consider how large can the backreaction from the scalar field to the metric be. For this, we assume that the product $\phi_c \beta$ is small, such that at the lowest order 
in $\alpha^{-1}$, the solutions are close to GR, and we ask whether the first order corrections can cause large 
deviations or not. We assume that the large $\alpha$ expansion holds 
for the scalar field, i.e. we must have  $\alpha{^{-1} }\phi_1 < \phi_c$ everywhere 
inside the star. The question we ask is: how large are the corrections to the metric when the
large $\alpha$ approximation for the scalar field approaches its limit of validity? There are two ways 
to approach such a limit; one is to make $\La$ larger so that the 
profiles for $\Phi_1$ are amplified, and the other is to take
an initial condition $\Phi_1(r_{min})$ close to $\phi_c$. We explore both possibilities to answer the previous question.

\begin{figure}
\includegraphics[width=0.9\textwidth]{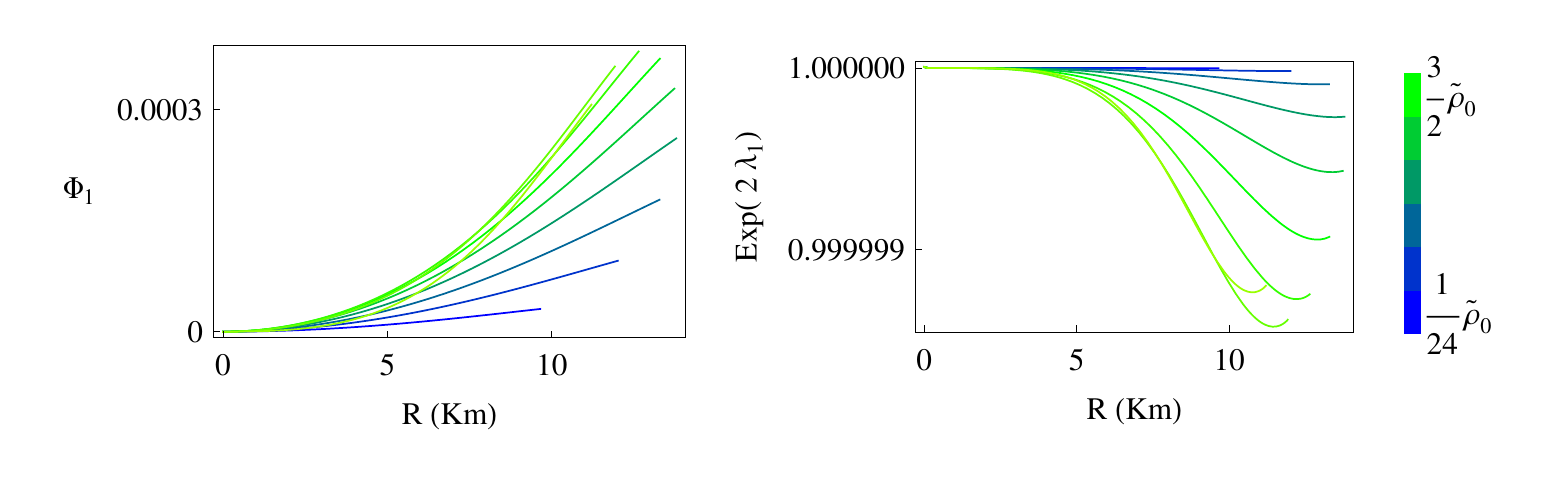} 
\caption{The left panel shows some scalar field profiles in the extreme case when $\Lambda$ is increased up to the limit when $\Phi_1(R)$ reaches values comparable to $\phi_c$, which is fixed 
to be $\phi_c=0.0005$. Even in this case, the corrections to $g_{rr}$ are still suppressed, as can be seen in the right panel. }\label{fig:corrections}
\end{figure}

In Fig.~\ref{fig:corrections}, we show $\Phi_1$ and the 
respective corrections to $g_{rr}$ for $\Lambda^3/\al \sim (10^{-11} {\rm  Km} )^{-3} $. The suppression of the large $\al$ is not valid any more, as can be seen from the fact that $\Phi_1$ becomes comparable to $\phi_c$ near the radius of the star.  Despite this, the effects on the metric, characterized by $\lambda_1$ in \eqref{neumet} are minimal.
We do not need to go further to conclude that this corrections do not alter the results displayed above for the binding energy and the mass-radius curves.

\begin{figure}
\includegraphics[width=0.9\textwidth]{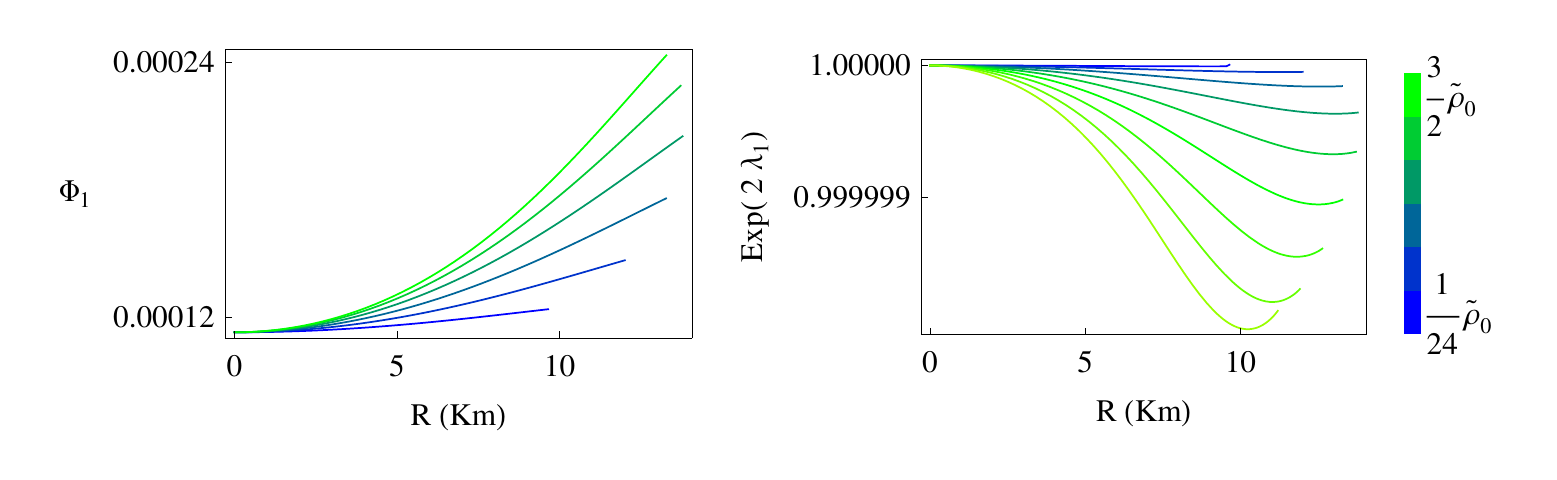}
\caption{The left panel shows some scalar field profiles in the extreme case when the initial condition
for $\Phi_1$ is increased up to values comparable to $\phi_c$, which is fixed 
to $\phi_c=0.0005$. As in Fig.~\ref{fig:corrections} the corrections to $g_{rr}$ are still suppressed, as can be seen in the right panel.}
\label{pf}
\end{figure}

As we mentioned before, the other possibility to approach the limit where the large-$\alpha$ approximation stops being reliable, is
to increase the value of the
initial conditions for $\Phi_1$, taking it closer to $\phi_c$. The results are shown in Fig.~\ref{pf} for $\Phi_1(r_{min})=0.0001$, while $\phi_c=0.0005$.  The deviations from GR are again negligible, although they are larger than the deviations caused by a large $\Lambda$. This can be understood as a result of $\Phi_1$ being of order $\phi_c$ all the way across the star, while for the $\Lambda$-induced corrections, $\Phi_1$ 
becomes large only near the surface of the star.  

Until now, everything seems to indicate that a gravitational configuration with a Galileon scalar field is hardly distinguishable from a pure GR solution. The $\mathcal O(\alpha^0)$ term of the
scalar field can be arbitrarily small for low densities and, under a reasonable physical requirement, we showed that it tends to decrease even more for higher densities. The Vainshtein radius derived from the first order corrections to the scalar field is in a good agreement with the result in the weak gravity limit,
and the effects of the scalar field $\Phi_1$ on the metric are highly suppressed. However, there is still an issue with the solutions for the scalar field beyond the critical density. We comment on this in the next subsection. 

%%%%%%%%%%%%%%%%%%%%%%%%%%%%%%%%%%%%%%%%%%%%%%%%%%%
\subsection{Critical density for the scalar field}
\begin{figure}
\includegraphics[width=0.7\textwidth]{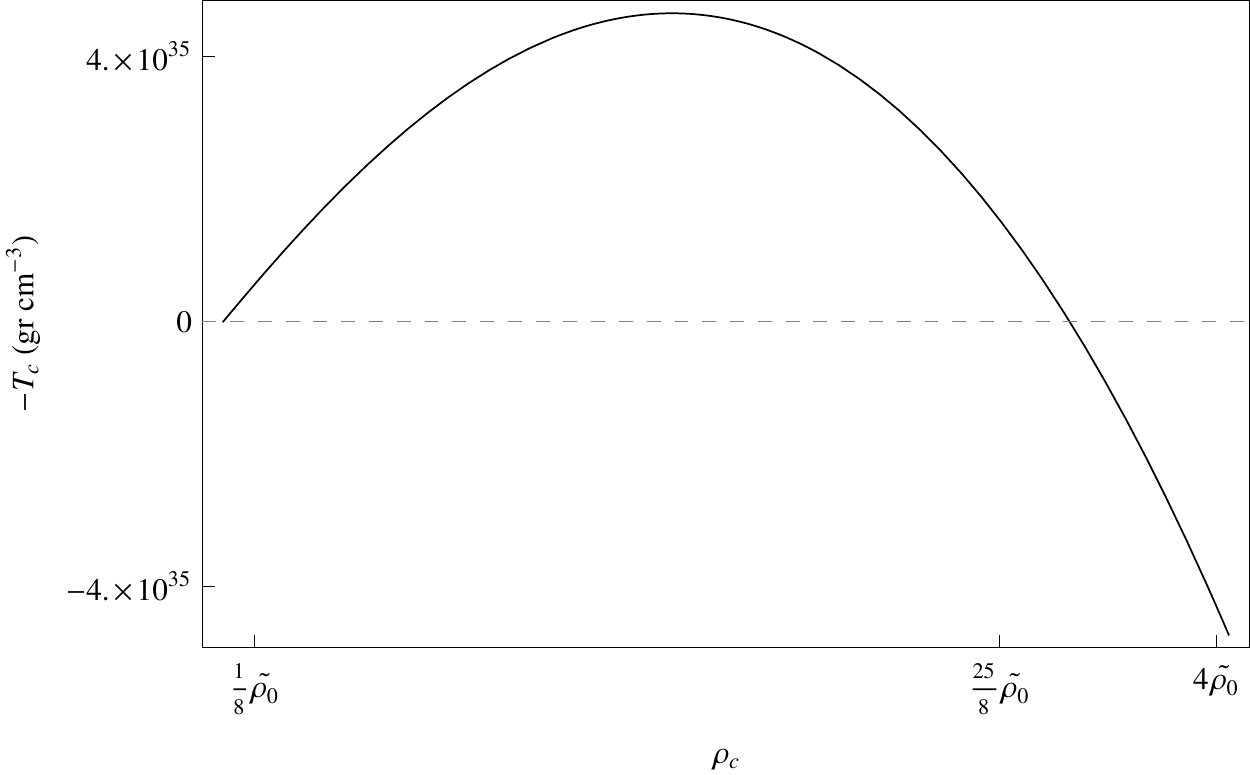} \caption{Central value of $T$ as a function of the central density. }\label{minusT}
\end{figure}

%%%%%%%%%%%%%%%%%%%%%%%%%%%%%%%%%%%%%%%%%%%%%%%%%%%
For general static solutions in vacuum, one of the beauties of the equation of motion for the Galileon scalar field is that it can be integrated once and it can be written as a simple quadratic algebraic equation for $\Phi'$, \eqref{zeta}. 
In the presence of matter, this cannot be done as the density and pressure are unknown functions of $r$. However, it turns out that the quadratic nature of the scalar field equation eventually shows up. The value of the density for which the scalar field solution ceases to exist is not the density corresponding to the maximum mass for neutron stars, but the density
at which the trace of the energy momentum tensor changes sign, causing the scalar field to become complex. 
A reason for this behaviour can be inferred from the the scalar field solution  \eqref{tmscalarfield}, where $\tilde \beta$ plays the same
role as the trace of the energy-momentum tensor, $T$. The parameter $\tilde \beta$ appears inside the square root, and if it changes sign at some radius, the solution ceases to exist. This is what happens inside neutron stars when $T$ changes sign. Fig.~\ref{minusT} shows the dependence of $T$ on the density at $r_{min}$. The existence of a critical density, $\rho_{crit}$, where $T$ changes sign is generic for relativistic matter, as we will explain further below.

\begin{figure} \includegraphics[width=0.7\textwidth]{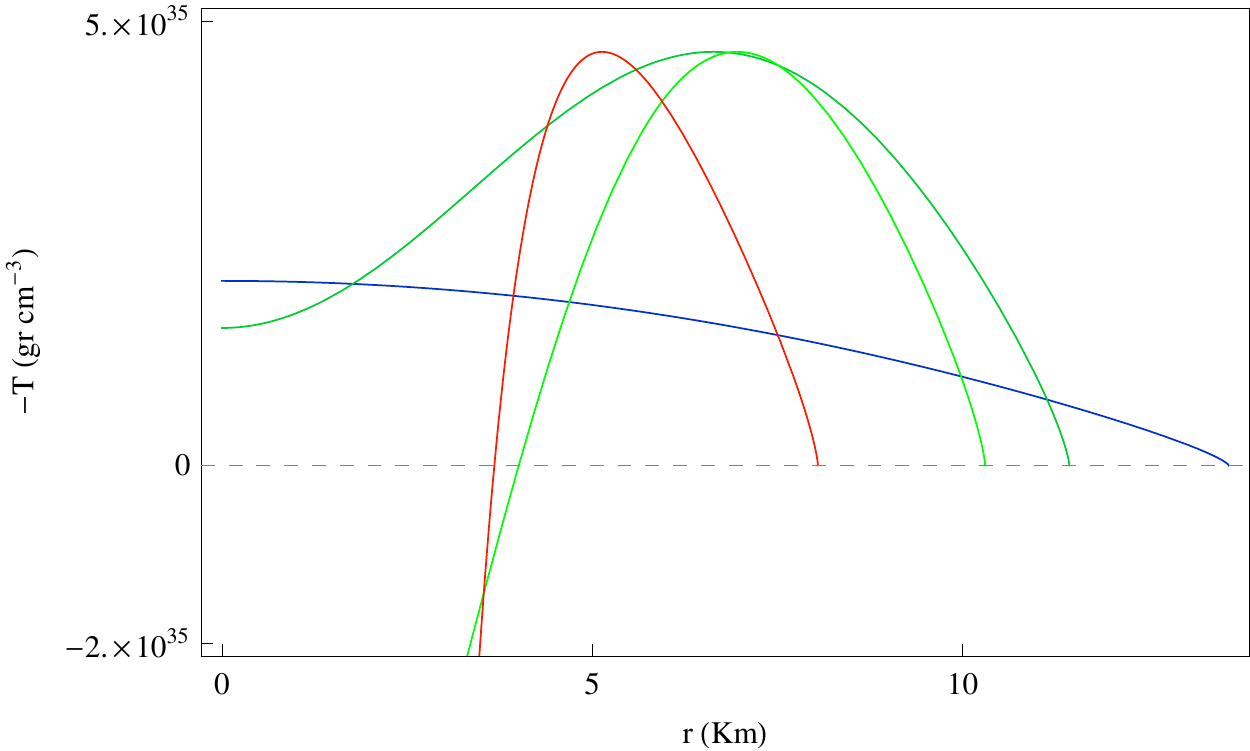}\caption{Profiles of $-T=\rho-3 p$ for two stars with $\rho_c<\rho_{crit}$  (blue and dark green curves) and other two with $\rho_c>\rho_{crit}$. For the stars with $\rho_c$ larger the critical density, $-T$ is negative at $r_{min}$ and it changes sign inside the star again at $r>r_{min}$. }\label{minusTvr}
\end{figure}

A naive way around would be to change the sign of the 
scalar field coupling constant in the hope of finding a branch of solutions for $\rho>\rho_{crit}$. This
certainly works in the flat space toy model, as can be seen from equation \eqref{tmscalarfield}.   
However, as shown in Fig.~\ref{minusTvr}, for the configurations
beyond $\rho_{crit}$, the trace of the energy momentum tensor shows a further change of sign inside the star (this is independent of the sign of $\beta$), thus a continuous solution for the scalar field does not seem to exist. 

\begin{figure}
\includegraphics[width=1.0\textwidth]{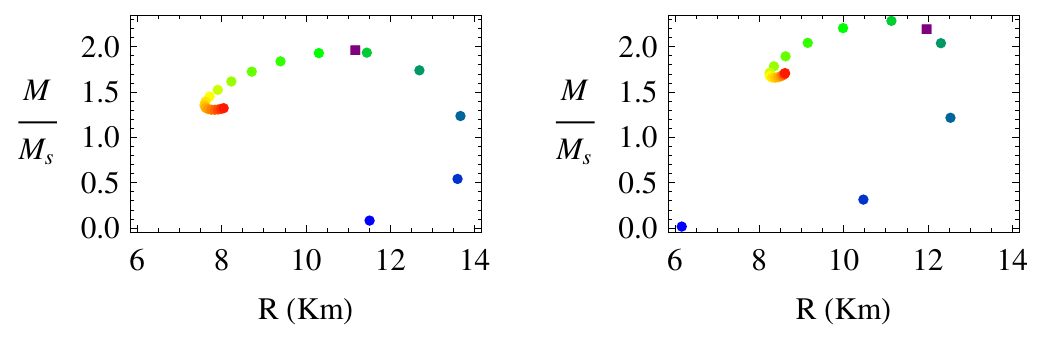}\caption{
Mass-radius curves for $\Gamma=2.34$ (left) and $\Gamma=2.84$ (right). $\Gamma=2.34$ is the value used in the rest of the calculations
through this work. The square mark indicates the configuration for which $T_c$ changes sign, which does not necessarily coincide with the maximum mass.}\label{rhocrit}
\end{figure}

The exact value of $\rho_{crit}$ depends upon the particular polytropic exponent and polytropic constant for the stellar model, and it can be larger or smaller than the density corresponding to the maximum mass.  The maximum mass of a neutron star is still uncertain both from  
a theoretical and an observational point of view (see e.g., \cite{Chamel2013} for a review). However, it is known for sure that it cannot be less than $1.97\pm0.04 M_\odot$, which corresponds to the mass of the pulsar J$1614-2230$ \cite{20981094}.  The particular polytropic parameters that we are using are adjusted in such a way that the predicted maximum mass is in agreement with this last value. Coincidentally,  the onset of radial instabilities, signalled by the maximum mass and the change of sign for $T$, happen roughly at the same density, as can be seen in the left panel of Fig.~\ref{rhocrit}. One may ask what happens if the maximum mass of neutron stars turns out to be larger than the current observational value? Comparing the left and right panel of Fig.~\ref{rhocrit}, we learn that as the maximum mass increases by changing the polytropic parameters, the critical density for the existence of the scalar field decreases.
This allows us to state that for a neutron star described by a polytropic equation of state, which is in agreement with the (current) observational maximum mass, the configurations that would collapse and form a black hole cannot support a non-trivial scalar field. 

The (non-)existence of high density stars in modified gravity models has been discussed in the literature extensively. In the context of massive gravity, it was first claimed that there is no non-singular solution that connects the asymptotic solutions to the solutions inside a source \cite{Damour:2002gp}. Later Ref.~\cite{Babichev:2009jt, Babichev:2010jd} found that this was a numerical artefact and they constructed numerically and analytically a solution that connects smoothly the solution inside a star to the asymptotic flat spacetime. However, it was found that, when the density of the object increases, the numerics becomes unstable and singularities are found to appear. It is still not clear whether those singularities are physical or just numerical artefacts. The existence of neutron stars was also debated in $f(R)$ gravity. It was claimed that there is an upper bound for the mass of neutron stars as the solution could not be found for high density stars \cite{Frolov:2008uf, Kobayashi:2008tq}. However, 
it was shown later that this was also due to numerical artefacts \cite{Babichev:2009fi, Babichev:2009td, Upadhye:2009kt}. At the same time, it was found that static solutions can be found only when $P < \rho/3$ \cite{Babichev:2009fi, Babichev:2009td}. This is closely related to our finding although the reason for the non-existence of the static solutions is quite different. In $f(R)$ gravity case, there is a potential for the scalar field. The coupling to the trace of energy-momentum tensor leads to the density (and pressure) dependent effective potential. When $P < \rho/3$, the effective potential does not have a minimum and stable static solutions cannot exist. In our case, there is no potential. Instead, the non-liner derivative coupling is responsible for suppressing the scalar field. We argued that the sign change of $-T = \rho-3P$ makes the solution complex and the solution ceases to exist. We note that a similar problem was found even in weak gravity when the density becomes negative in voids \cite{
Barreira:2013eea}. 


\section{Conclusions}
In this paper, we developed an expansion scheme to solve the non-linear equations inside the Vainshtein radius in a system with the cubic Galileon term. If applied to a test scalar field in the Minkowski spacetime, this expansion is controlled by $(r/r_V)^{3/2}$, which is the same expansion obtained in a dual theory introduced in \cite{Gabadadze2012, Padilla:2012ry}. We applied this expansion method and found the first order corrections to the metric and the scalar field due to the Galileon term for spherically symmetric configurations. We also included a slow rotation in the system and derived again the first order correction. The integration constants in the vacuum solutions can be determined by introducing a star. In order to couple matter, we choose two different couplings, one linear and other queadratic in the scalar field. We start from the linearly coupled model and derived corrections to a low-density TOV solution. This solution could be connected smoothly to the exterior Schwarzschild solution with small corrections due to the Galileon term. We confirm that all deviations from the GR solutions are strongly suppressed within the Vanshtein radius, including corrections to the rotation. 

In the quadratically coupled models, in which ``scalarisation'' can strongly enhance deviations from GR in standard scalar tensor gravity, we showed that the Vainshtein mechanism suppressed the difference between the scalar field at the centre of the star and its asymptotic value, hence the scalarisation effect does not take place. Finally, we found, numerically, the corrections to relativistic star solutions due to the Galileon term. As expected, the higher the density, deviations from GR in the star are more suppressed for a given asymptotic value of the scalar field. Although in the strong gravity regime, the Vainshtein radius does not grow as the density increases, unlike in the weak gravity regime, this is not enough to have detectable deviations from GR. One caveat of our approach is that our expansion scheme is valid only inside the Vainshtein radius while the asymptotic solution is valid outside the Vainshtein radius. Thus strictly speaking, our matching at the Vainshtein radius is not well justified. Using 
the exact solution for a scalar field in the Minkowski spacetime, we checked that the inaccuracies caused by this procedure does not change orders of magnitude. Still we cannot guarantee that the matching always exists for the full solution. Full numerical studies of the solution are left for future investigations. Our approximated solutions will be useful to find exact solutions as it is a challenging task to construct a relativistic star numerically due to the hierarchy of scales in the system (i.e. the size of the star and the Vainshtein radius). 

It is well know that these Vainshtein solutions do present, generically, sub- and superluminal propagation \cite{Nicolis:2008in}.  Our solution for the scalar field becomes the same as the flat spacetime solution at large radii $r \gg r_s$. Thus we expect that we cannot avoid superluminality. The situation may differ near the Schwarzschild radius due to the effects of the background curvature. This requires us to study time dependent fluctuations around our static solutions. This is an interesting problem and we leave this for future work. 

Finally, we found that the existence of the scalar field solution is not always guaranteed in the strong gravity regime. When the trace of the energy momentum tensor $-T = \rho - 3 P$ changes sign, the scalar field solution ceases to exist. We argued that this was due to the quadratic nature of the scalar field equation and the solution became complex at this point. For the equation of state that we used in this paper, this happens close to the maximum mass that corresponds to the onset of the gravitational instability. This equation of state is adjusted to explain the observed maximum mass of neutron stars. This indicates that unstable neutron stars cannot carry a scalar charge although we cannot exclude a possibility of dynamical solutions or non-spherically symmetric solutions with a non-trivial scalar configuration. By changing the equation state, the maximum mass does not coincide with the sign change of the trace of the energy-momentum tensor. However, we found that if the maximum mass increases, the 
critical density, above which the scalar field solution ceases to exist, decreases. This implies that, for a neutron star described by a polytropic equation of state in agreement with the observational maximum mass, a cubic Galileon scalar field cannot exist for the configurations that would collapse and form a black hole. Along the same line, there is a no-hair theorem for black holes in the presence of the Galileon scalar field \cite{Hui2013}. This theorem proves that static, spherically symmetric black hole solutions cannot sustain non-trivial scalar profiles. It is still an open question what is the fate of the  scalar field when a star's density increases dynamically and eventually collapses to become a black hole. 

\acknowledgments
JC acknowledges support through CONACYT grants 290649 and 290749. KK is supported by the UK Science and Technology Facilities Council (STFC) grants ST/K00090/1 and ST/L005573/1. GN is supported by the grant CONACYT/179208. 
GT thanks STFC for financial support through the grant ST/H005498/1.

\bibliographystyle{plain}

 \end{document}